\documentclass{emulateapj}

\usepackage{color}
\newcommand{\colr}{\textcolor{black}}

\shorttitle{Low-frequency solar p modes in spatially-resolved helioseismic data}
\shortauthors{Salabert et al.}

\begin{document}

\title{Measurement of low signal-to-noise-ratio solar p modes \\ in spatially-resolved helioseismic data}

\author{D.~Salabert\altaffilmark{1,*}, 
       J.~Leibacher\altaffilmark{1,2}, 
       T.~Appourchaux\altaffilmark{2}, 
       and F.~Hill\altaffilmark{1}} 

\altaffiltext{1}{National Solar Observatory, 950 North Cherry Avenue, Tucson, AZ 85719}
\altaffiltext{*}{Now at: Instituto de Astrof\'isica de Canarias, C/ V\'ia L\'actea s/n, 38205 La Laguna, Tenerife, Spain, salabert@iac.es}
\altaffiltext{2}{Institut d'Astrophysique Spatiale, CNRS-Universit\'e Paris XI UMR 8617, 
      91405 Orsay Cedex, France}

\begin{abstract}
We present an adaptation of the rotation-corrected, $m$-averaged spectrum technique designed to observe low signal-to-noise-ratio, low-frequency solar p modes. The frequency shift of each of the $2l+1$ $m$ spectra of a given ($n,l$) multiplet is chosen that \colr{maximizes the likelihood} of the $m$-averaged spectrum. A high signal-to-noise ratio can result from combining individual low signal-to-noise-ratio, individual-$m$ spectra, none of which would yield a strong enough peak to measure. We apply the technique to GONG and MDI data and show that it allows us to measure modes with lower frequencies than those obtained with classic peak-fitting analysis of the individual-$m$ spectra. 
We measure their central frequencies, splittings, asymmetries, lifetimes, and amplitudes. 
The low-frequency, low- and intermediate-angular degrees rendered accessible by this new method 
correspond to modes that are sensitive to the deep solar interior down to the core ($l \leq 3$) 
and to the radiative interior ($4 \leq l \leq 35$). Moreover, the low-frequency modes have deeper 
upper turning points, and are thus less sensitive to the turbulence and magnetic fields of the outer 
layers, as well as uncertainties in the nature of the external boundary condition. As a result of 
their longer lifetimes (narrower linewidths) 
at the same signal-to-noise ratio the determination of the frequencies 
of lower-frequency modes is more accurate, and the resulting inversions should be more precise.
\end{abstract}


\section{Introduction}
Our knowledge of the structure and dynamics of the solar interior has been considerably improved by the use of measurements of the properties of the normal modes of oscillation of the Sun. However, the Sun's interior is far from being fully understood, and better measurements of the mode parameters will also help to better understand the mode excitation and damping mechanisms as well as the physical properties of the outer layers by better constraining the turbulence models. A large number of predicted acoustic oscillation modes, defined by their radial orders ($n$) and their angular degrees ($l$), are not yet observed in the low-frequency range (i.e., approximately below 1800~$\mu$Hz), because the amplitude of the acoustic modes decreases as the mode inertia increases as the frequency decreases, while the solar noise from incoherent, convective motions increases: thus the signal-to-noise ratio (SNR) of those modes is progressively reduced. Moreover, these low-frequency p modes have very long lifetimes, as much as several years, which results in very narrow linewidths, hence precise frequency measurements. Thanks to the long-duration helioseismic observations collected by the spaced-based instruments Michelson Doppler Imager (MDI) \citep{scherrer95} and Global Oscillations at Low Frequencies (GOLF) \citep{gabriel95} onboard the Solar and Heliospheric Observatory (SOHO) spacecraft, and by the ground-based, multi-site Global Oscillation Network Group (GONG) \citep{harvey96} and Birmingham Solar Oscillations Network (BiSON) \citep{chaplin96}, the frequency resolution is continuously improving and the observation of lower radial-order solar p modes is becoming possible. Their precise mode parameter determination is of great interest for improving our resolution throughout the solar interior because they cover a broad range of horizontal phase velocity, and thus a broad range of depths of penetration. Moreover, these low-frequency modes have lower reflection points in the outer part of the Sun, which make them less sensitive to the turbulence and the magnetic fields in the outer layers, where the physics is poorly understood.

The usual mode-fitting analysis consists of fitting the $2l+1$ individual-$m$ spectra of a given multiplet ($n,l$), either individually or simultaneously. Such fitting methods fail to obtain reliable estimates of the mode parameters when the SNR \colr{of the individual-$m$ spectra} is low. Instead, various pattern-recognition techniques have been developed in an effort to reveal the presence of modes in the low-frequency range \citep[see e.g.,][and references therein]{schou98,app00,chaplin02,broomhall07}. In the case of spatially-resolved helioseismic data (such as GONG and MDI observations), $m$-averaged spectra appeared to be a powerful tool, since for a given multiplet ($n,l$), there exist $2l+1$ individual-$m$ spectra, which can result in an average spectrum with a SNR~$\gg 1$ once the individual-$m$ spectra are corrected for the rotation- and structure-induced \colr{frequency} shifts. The $m$-averaged spectra were employed early in the development of helioseismology by \citet{brown85}, but were replaced by fitting the $m$ spectra individually as the quality and the SNR of the data improved. However, years later, in order to fully take advantage of the long-duration helioseismic GONG and MDI instruments and reach lower frequencies in the solar oscillation spectrum, \citet{schou98,schou02,schou04} and \citet{app00} used the $m$-averaged spectra corrected by the modeled solar rotation to detect new low radial-order p modes and to set upper limits on the detectability of the g modes. These authors demonstrated the potential advantage of such rotation-corrected, $m$-averaged spectra. 

We present here an adaptation of the $m$-averaged spectrum technique in which the $m$-dependent shift parameters are determined by maximizing the quality of the resulting average spectrum. The analysis is performed on long-duration time series of the spatially-resolved helioseismic GONG and MDI observations of the low- and medium-angular degrees ($1 \leq l \leq 35$). This range of oscillation multiplets samples the radiative interior down to the solar core. In Sec.~\ref{sec:obs}, we introduce the different datasets used in this analysis. In Sec.~\ref{sec:method}, we describe this new technique in order to observe low signal-to-noise-ratio, low-frequency p modes, explaining the different steps of the analysis from the mode detection to peak-fitting. In Sec.~\ref{sec:compa}, we demonstrate that this method allows us to successfully measure lower-frequency modes than those obtained from classic peak-fitting analysis of the individual-$m$ spectra by comparing with other measurements obtained from coeval datasets. In Sec.~\ref{sec:data_results}, we present the mode parameters of these long-lived, low-frequency acoustic modes down to $\approx$~850~$\mu$Hz extracted from the analysis of 3960 days of GONG observations using the $m$-averaged spectrum technique. Finally, Sec.~\ref{sec:conc} summarizes our conclusions.

\section{Observations}
\label{sec:obs}
Details of the spatially-resolved helioseismic observations collected by both GONG 
and MDI used for this work (the starting and ending dates, and their corresponding 
duty cycles) are given in Table~\ref{table:series}. Coeval 2088-day observations of 
GONG and MDI were analyzed for oscillation multiplets with angular degrees from $l=1$ 
to $l=35$, and are then directly compared to those of \citet{korz05} for $l \leq 25$ 
measurements of the same datasets. We also applied the analysis to 3960 days of GONG 
data ($1 \leq l \leq 35$), which constitutes so far the longest time 
series ($\approx$~11 years, spanning most of solar cycle 23) of spatially-resolved 
observations analyzed. 

\begin{table}
\begin{center}
\caption{Details of the long GONG and MDI analyzed time series}
\begin{tabular}{lllc}
\hline\hline
      & \multicolumn{1}{c}{Start} & \multicolumn{1}{c}{End}  & \multicolumn{1}{c}{Fill}  \\ 
\multicolumn{1}{c}{Series} & \multicolumn{1}{c}{Date}  & \multicolumn{1}{c}{Date} & \multicolumn{1}{c}{(\%)}  \\ \hline     
3960-day GONG & 1995 May 7 & 2006 Mar 9  &  84.6  \\
2088-day GONG & 1996 May 1 & 2002 Jan 17 &  83.7  \\
2088-day MDI  & 1996 May 1 & 2002 Jan 17 &  88.9  \\
\hline
\end{tabular}
\label{table:series}
\end{center}
\end{table}

\section{Method}
\label{sec:method}
An $m$-averaged spectrum corresponds to the average of the $2l+1$ individual-$m$ components of an oscillation multiplet ($n,l$), thus reducing the non-coherent noise. Before averaging, each $m$ spectrum of a given mode ($n, l$) is shifted by a frequency that compensates for the effect of differential rotation and structural effects on the frequencies. The $m$-averaged spectrum concentrates, for a given multiplet ($n, l$), all of the $2l+1$ $m$ components, as it would be if the Sun were a purely-spherical, non-rotating object. Thus, the average of the $2l+1$ individual-$m$ spectra considerably improves the SNR of the resulting $m$-averaged spectrum.

\begin{figure*}[t]
\centering
\includegraphics{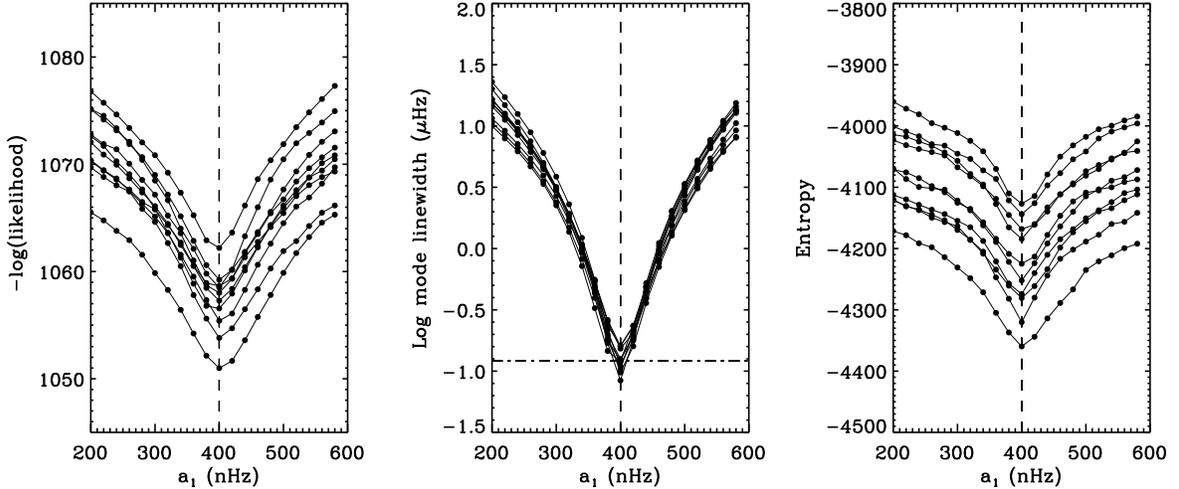}
\caption{\colr{$- \log(\rm{likelihood})$} ({\it left panel}), natural logarithm of the mode linewidth ({\it middle panel}), 
and entropy ({\it right panel}) of the $m$-averaged spectrum as a function of the shift 
coefficient~$a_1$ obtained from ten simulated $l=15$ spectra. The vertical dashed lines 
represented the introduced value of $a_1$ (400~nHz) in the simulations. The horizontal 
dotted-dashed line in the center panel represents the mode linewidth of the simulated spectra, 0.4~$\mu$Hz.}
\label{fig:fom}
\end{figure*}

\subsection{Determination of the shifts}
\label{ssec:det_acoef}
The $m$-averaged spectrum is obtained by finding the estimates of the splitting coefficients, 
commonly called $a$-coefficients, which \colr{maximizes the likelihood} of the $m$-averaged spectrum. 
The $a$-coefficients are individually estimated through an iterative process, with the initial 
values taken from a model. Thus, for a given mode ($n,l$), the frequency shift $\delta\nu_{n l m}$ 
is parameterized by a set of coefficients, as:

\begin{equation}
 \delta\nu_{n l m} = \sum_{i=1}^{i_{max}}{a_i(n,l) P^{i}_{l,m}},
\label{eq:poly}
\end{equation}

\noindent
where $a_i(n,l)$ are the splitting coefficients, and $P^i_{l,m}$ corresponds
to the Clebsch-Gordan polynomial expansion as defined by \citet{ritz91}. 
In this definition, the odd orders of the $a$-coefficients describe the effects 
of solar rotation, while the even orders correspond to departures from spherical 
symmetry in the solar structure as well as to quadratic effects of rotation. 
Each $a_i$ is chosen to \colr{maximize the likelihood} of the $m$-averaged 
spectrum. This is performed through an iterative procedure. For a particular order $i$ of 
the coefficients $a_i$, a range of values is scanned around its initial value, while the 
other $a_{i'\neq i}$s are kept fixed to their previously estimated values. 

For each scanned value of $a_i$, the individual-$m$ spectra are shifted by the 
corresponding Clebsch-Gordan polynomials, and the mean of these $2l+1$ shifted spectra 
is taken. The mean power spectrum is then fitted using a Maximum-Likelihood Estimator (MLE) 
minimization as described in Sec~\ref{ssec:extraction} and its likelihood determined. 

For a Monte-Carlo simulation, the left panel of Fig.~\ref{fig:fom} shows the variation of 
the likelihood from the MLE minimization as a function of the first splitting coefficient $a_1$, 
showing a well defined minimum which represents the best value of $a_1$. The artificial power 
spectra were simulated following the methodology described in \citet{fierry98}. We also examined 
the sensitivity of the mode linewidth and the entropy as criteria for determining the best shifts. 
\colr{In our case, the entropy \citep{shannon48} can be seen as a measure of randomness 
in the $m$-averaged spectrum, $\mathcal{S}$, and is defined as $-\sum [\mathcal{S} \times \ln \mathcal{S}]$.}

Both linewidth and entropy show well 
defined minima around the input value of $a_1$ (middle and right panels of Fig.~\ref{fig:fom} 
respectively). Indeed, the $m$-averaged spectrum gets narrower as $a_1$ converges to its input 
values of 0.4~$\mu$Hz and $a_1 = 400$~nHz. Similar variations are obtained for all the $a_i$s. 
As detailed in Appendix~\ref{sec:foms}, the use of these different criteria to determine
the best estimates of the $a$-coeficients returns consistent results.  

The iteration is performed until the difference between two iterations in each of the 
computed $a_i$ coefficients falls below a given threshold (such as 0.25$\sigma$ in the 
case of $a_1$). Also, \colr{in order to remove any outliers}, some quality checks are 
performed after each measure of an $a_i$ which needs to fall within a constrained range of values. 
\colr{For example, a $\pm$15\% window around its theoretical expectation is used for $a_1$.} 
Here, we fitted only the six first $a_i$ in the Clebsch-Gordan expansion, even though the quality of the data supports the determination of higher-order coefficients. 

\begin{figure*}[t]
\centering
\includegraphics[width=\textwidth]{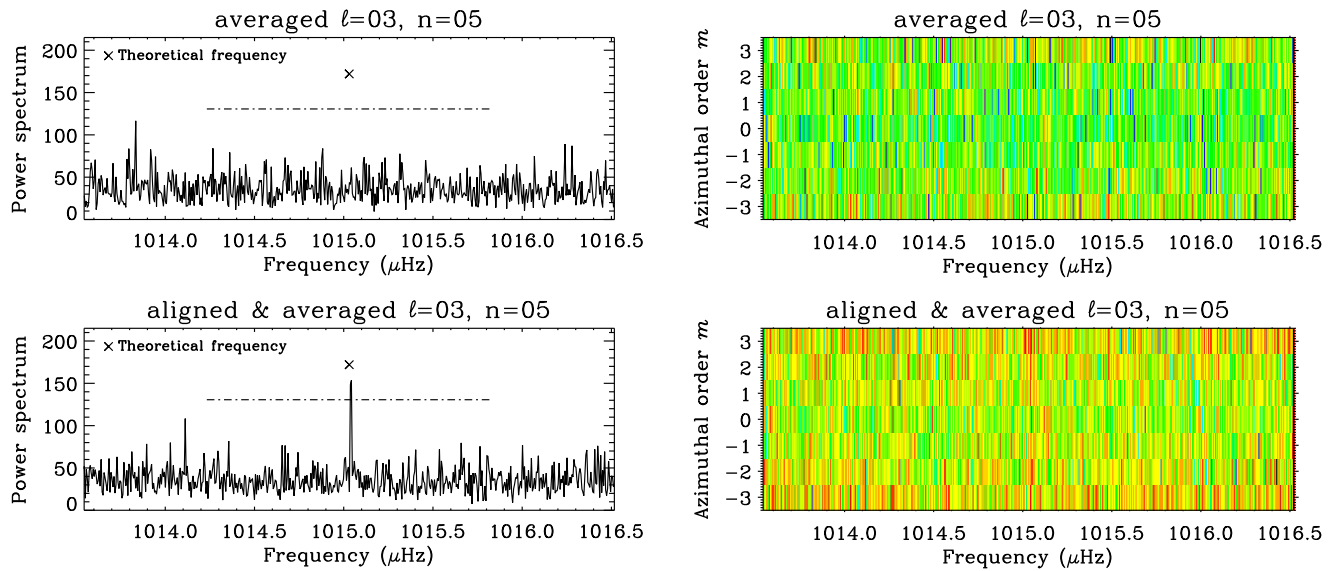}
\caption{Examples of $m$-averaged power spectra ({\it left panels}) before ({\it top}) and 
after ({\it bottom}) correcting for the shift coefficients in the case of the oscillation 
multiplet $l=3$, $n=5$ observed in 2088 days of GONG data. The corresponding $m-\nu$ diagrams 
are also shown ({\it right panels}). The crosses indicate the position of the corresponding 
theoretical central frequency calculated from Christensen-Dalsgaard's model~S \citep{jcd96}. 
The dot-dashed lines on the left hand-side panels give the 10\% probability limit that a peak 
is due to noise in \colr{the 238 windows, 1.5~$\mu$Hz wide}. The \colr{illustrated} spectral window \colr{contains} the $2l+1$ components of the \colr{represented} multiplet.}
\label{fig:spec_l3}
\end{figure*}

\begin{figure*}[t]
\centering
\includegraphics[width=\textwidth]{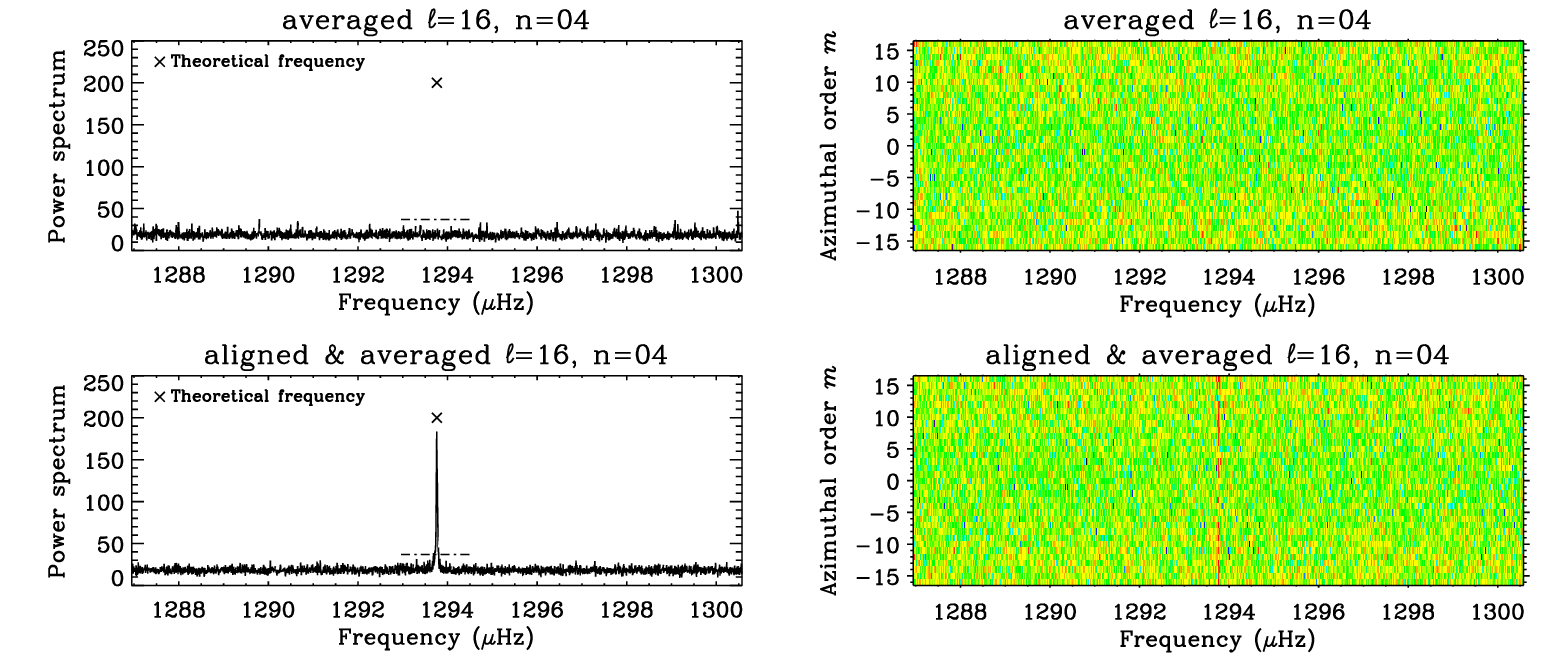}
\caption{Examples of $m$-averaged power spectra ({\it left panels}) before ({\it top}) 
and after ({\it bottom}) correcting for the shift coefficients in the case of the oscillation 
multiplet $l=16$, $n=4$ observed in 2088 days of GONG data. The 
corresponding $m-\nu$ diagrams are also shown ({\it right panels}). The crosses indicate 
the position of the corresponding theoretical central frequency calculated 
from Christensen-Dalsgaard's model~S \citep{jcd96}. The dot-dashed lines on the left 
hand-side panels give the 10\% probability limit that a peak is due to noise in 
\colr{the 238 windows, 1.5~$\mu$Hz wide}. The \colr{illustrated} spectral window \colr{contains}  the $2l+1$ components of the \colr{represented} multiplet.}
\label{fig:spec_l16}
\end{figure*}

\begin{figure*}[t]
\centering
\includegraphics[width=\textwidth]{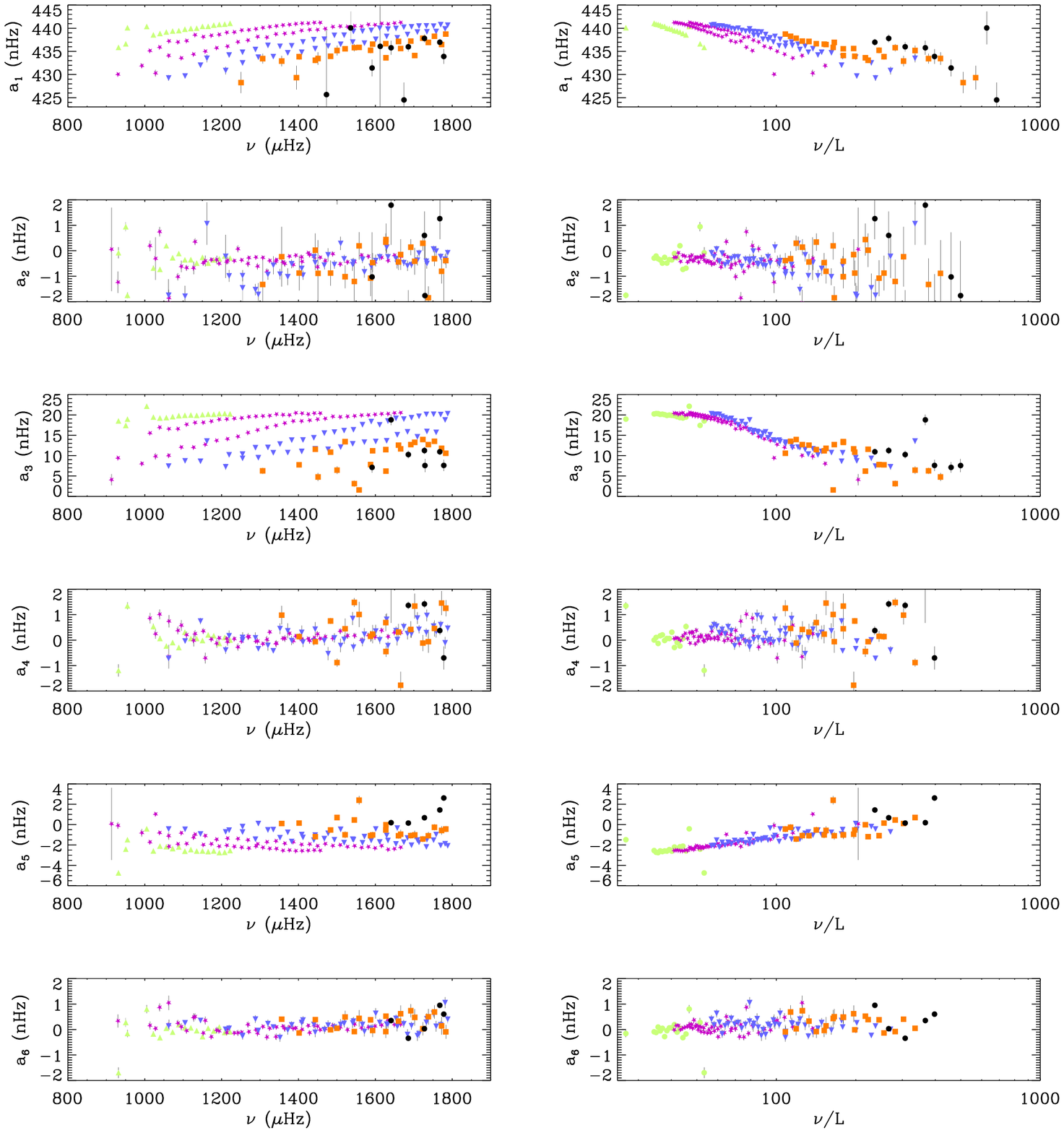}
\caption{First six $a_i$ splitting coefficients of the low-frequency p modes 
with $1\leq l \leq 35$ measured in
3960 days of GONG data. The $a$-coefficients are represented
as a function of the mode frequency $\nu$ ({\it left column}), 
and $\nu/L$, with $L=\sqrt{l(l+1)}$ ({\it right column}). 
The different colors and symbols correspond to selected ranges of radial 
orders $n$: green triangles, modes with $n=1,2$; purple stars, $n=3,4$; 
blue upside-down triangles, $n=5,6$; orange squares, $n=7,8$; and black dots, $n\geq9$.} 
\label{fig:acoefs3960d}
\end{figure*}

\colr{Finally, low SNR peaks in the $m$-averaged spectrum (after adjustment) are tested against the
H0 hypothesis.  In the framework of that hypothesis, the resulting spectra are tested against a statistics
pertaining to pure noise ($\chi^2$ with 2(2$l$+1) d.o.f).  This test has been widely applied to 
helioseismic observations in the search for long-lived, low radial-order p modes 
and g modes \citep[see e.g.,][]{app00}.}  In the present analysis, we rejected peaks that have 
a greater than 10\% chance of being \colr{due to noise} in the \colr{238} analyzed windows, \colr{each containing 288 frequency bins.  Here the fixed number of bins was chosen because we know that the range of theoretical frequency lie within 1.5~$\mu$Hz or so}.
Figures~\ref{fig:spec_l3} and \ref{fig:spec_l16} illustrate the advantage of using the $m$-averaged 
spectrum technique in the case of two oscillation multiplets for 2088 days of GONG data, where 
the $m$-averaged spectra before and after the correction for the splitting coefficients are shown. 
These examples show the $m$-averaged spectra of the modes $l=3$, $n=5$ at $\approx$~1015.0~$\mu$Hz 
(Fig.~\ref{fig:spec_l3}), and $l=16$, $n=4$ at $\approx$~1293.8~$\mu$Hz (Fig.~\ref{fig:spec_l16}), 
as well as the corresponding $m-\nu$ diagrams. These two examples were chosen to demonstrate the 
performance at different SNR levels. The corresponding 10\% probability levels are given. 
The $m-\nu$ diagrams in the case of the mode $l=3$, $n=5$ (right panels on Fig.~\ref{fig:spec_l3}) do 
not show any high SNR structure before or after correction.  However, the $m$-averaged spectrum after 
correction clearly shows the target mode (lower left-panel on Fig.~\ref{fig:spec_l3}), with an unambiguous 
detection level. The mode $l=16$, $n=4$ presents a higher SNR (Fig.~\ref{fig:spec_l16}) and its $m-\nu$ 
diagram 
shows that the individual-$m$ spectra line up after correction (lower-right panel on Fig.~\ref{fig:spec_l16}). 
The estimated splitting coefficients of the low-frequency modes with $1\leq l \leq35$ measured in 
the 3960-day GONG dataset are shown in Fig.~\ref{fig:acoefs3960d} as a function of frequency 
and $\nu/L$ (with $L=\sqrt{l(l+1)}$), which is approximately proportional to the sound speed at 
the mode's inner turning point. Modes with selected ranges of radial orders are represented with different colors and symbols.

\subsection{Extraction of the mode parameters}
\label{ssec:extraction}
For a given mode $(n,l)$, the best estimates of the splitting coefficients 
determined as discussed in Sec~\ref{ssec:det_acoef} are used to calculate 
its $m$-averaged spectrum. When $N$ independent power spectra are averaged 
together, the statistics of the mean power spectrum corresponds to a $\chi^2$ 
with $2\times N$ degree-of-freedom (d.o.f.) statistics. \citet{app03} demonstrated 
that the mean of $2l+1$ independent power densities, which has a $\chi^2$ with more 
than 2 d.o.f. statistics, can be correctly fitted with a Maximum-Likelihood Estimator (MLE) 
minimization code developed for spectra following a $\chi^2$ with 2 d.o.f. statistics.
The asymmetric Lorentzian model of \citet{nigam98} was used to describe the $m$-averaged spectrum, as:

\begin{equation}
P_{n,l}(\nu) = H_{n,l} \frac{(1+\alpha_{n,l} x_{n,l})^2+\alpha_{n,l}^2}{1+x_{n,l}^2} + B_{n,l},
\label{eq:mlemodel}
\end{equation}

\noindent
where 

\begin{equation}
x_{n,l}=\frac{2(\nu-\nu_{n,l})}{\Gamma_{n,l}}.
\end{equation}

\noindent
Then, for a given mode ($n,l$), the central frequency, the Full-Width-at-Half-Maximum (\textsc{fwhm}), 
and the power height of the spectral density are respectively $\nu_{n,l}$, $\Gamma_{n,l}$, and $H_{n,l}$. 
The peak asymmetry is described by the parameter $\alpha_{n,l}$, while
$B_{n,l}$ represents an additive, constant background level in the fitted window. The first spatial 
leaks ($\delta l=0$, $\delta m=\pm2$), commonly called $m$-leaks, are also included in the fitting model 
and added to Eq.~\ref{eq:mlemodel}. The frequencies of the $m$-leaks are set from the central frequency 
of the target mode using the previously measured splitting coefficients (Sec.~\ref{ssec:det_acoef}). 
Their peak asymmetries are assumed to be the same as that of the target mode, while their \textsc{fwhm}s 
are a free parameter of the fit and different from the target mode. The amplitude of the $m$-leaks is 
specified to be a fixed fraction of the central peak, which was estimated from the leakage matrix 
developed especially for the GONG \citep{hill98} and MDI (J. Schou, private communication) data. 
The first spatial leaks in the $m$-averaged spectrum 
were determined by averaging for a given multiplet ($n,l$) the $\delta m \pm 2$ leaks over the entire $2l+1$ spectra.

The size of the fitting window, \colr{$\Omega_\nu$}, is proportional to the first estimates of the mode \colr{width, $\Gamma_{n,l}$,} and centered around the frequency of the target mode. 
\colr{It is defined as: 
\begin{equation}
\Omega_\nu = 20\sqrt{\Gamma_{n,l}^2 + \Delta\nu_r^2} + \Delta_{\delta m},
\end{equation}
where $\Delta\nu_r$ is the frequency resolution of the power spectrum. The first spatial leaks are always included 
in the fitting range by adding the offset $\Delta_{\delta m} = 800$~nHz. The multiplicative factor $20$ ensures a 
good sampling of the mode profile in the low-frequency range. A comparable 
definition of the fitting window was adopted by \citet{korz05}.}
Bad fits were removed based on a set of quality criteria based on the fitted mode parameters and associated 
uncertainties, \colr{such as, (1) the error of the mode frequency must be less than its mode width; 
(2) the SNR must be larger than 1; and (3) the mode width must be larger than the frequency resolution.}
A discussion on the impact of the fitting model (asymmetry, spatial leaks) on the extracted mode 
parameters used to describe 
the $m$-averaged spectrum can also be found in the Appendix~\ref{sec:impmodel}.

Figure~\ref{fig:mspec} shows examples of the $m$-averaged power spectra for four different radial 
orders $n$ of the multiplet $l=17$, and the corresponding best MLE fits, which included the mode 
asymmetry and the $\delta m \pm 2$ spatial leaks. The blending of the first $m$ leaks is particularly 
clear as the linewidths increase with increasing frequency.

\subsection{Mode parameter and $a$-coefficient uncertainties}
\label{ssec:error}
The mode parameter uncertainties are established in the usual manner by the inverse of 
the covariance matrix. However, because the $m$-averaged spectrum is fitted using a MLE 
minimization and, as explained in \citet{app03}, the formal uncertainties must be normalized 
by the square root of the number of averaged spectra, i.e., in our case, by $\sqrt{2l+1}$. 
But this a-posteriori error normalization is correct only if the $2l+1$ spectra of a 
given ($n,l$) mode have the same variance (or SNR). Since the condition of equal SNR 
among the $m$ spectra within a multiplet is not satisfied in our case, the uncertainties 
of the mode parameters have to be taken as a first approximation only. However, Monte-Carlo 
simulations show that this error normalization holds even in 
the case of $m$-dependent SNR (see Sec.~\ref{ssec:montecarlo}).

\begin{figure*}[t]
\begin{center}
\begin{tabular}{cccc} 
\resizebox{4.4cm}{!}{\includegraphics[scale=0.5]{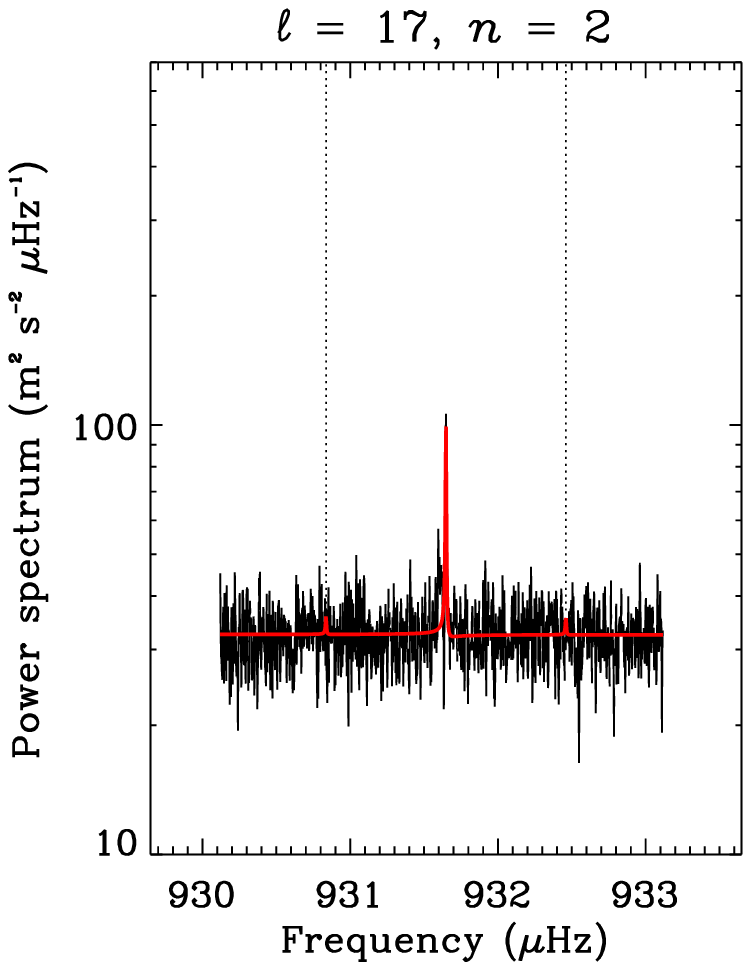}} &
\resizebox{4.4cm}{!}{\includegraphics[scale=0.5]{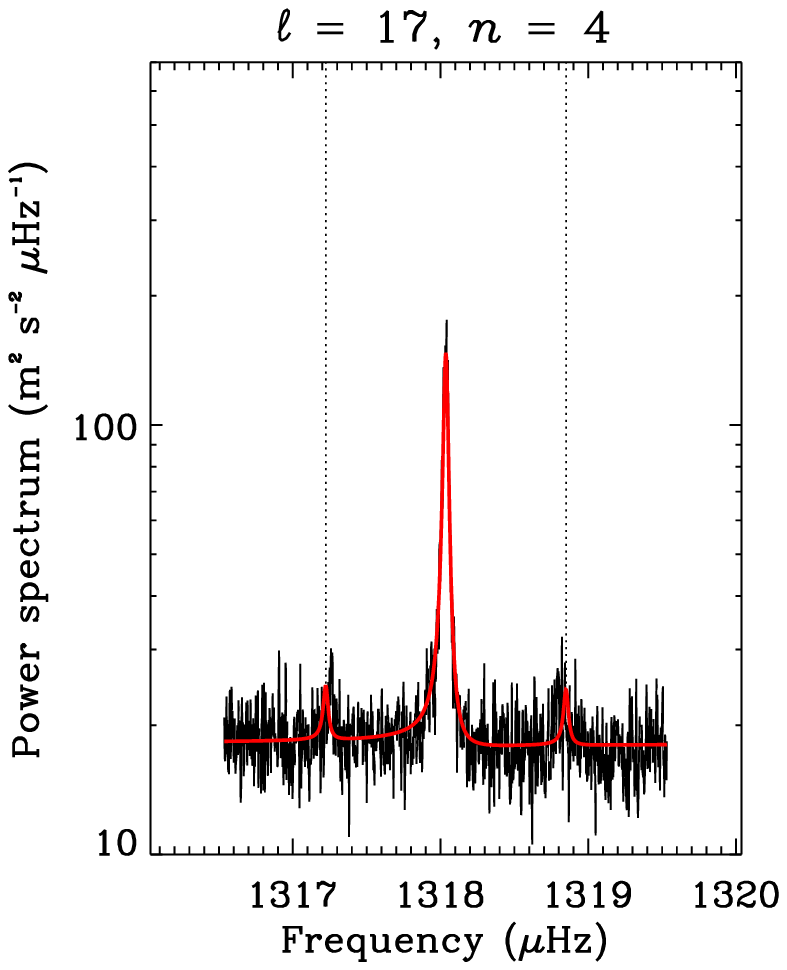}} &
\resizebox{4.4cm}{!}{\includegraphics[scale=0.5]{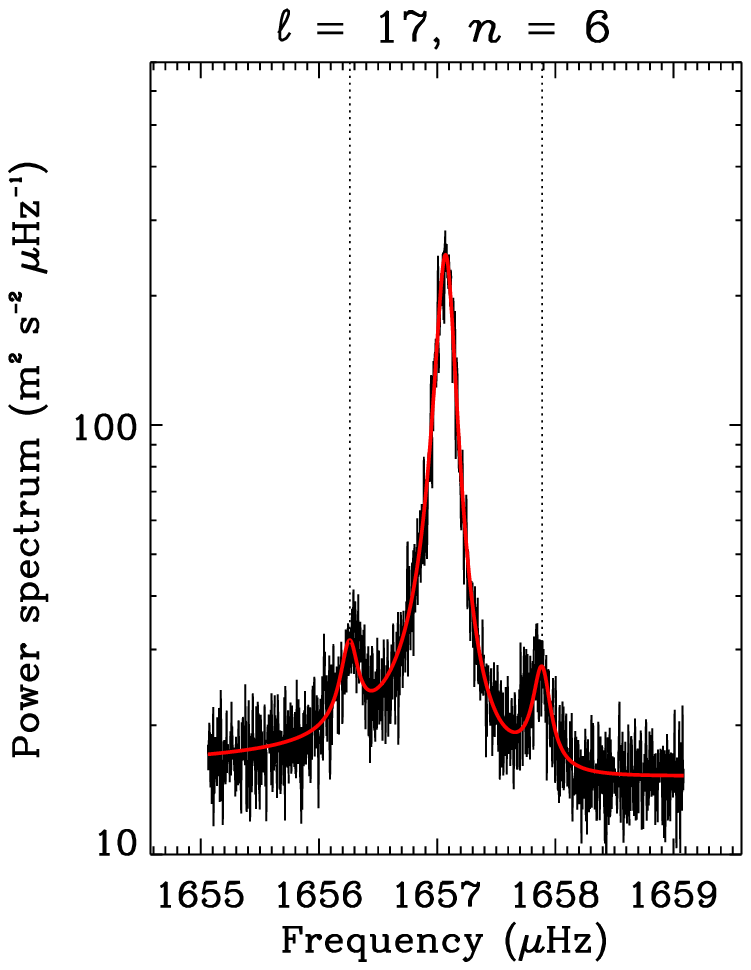}} &
\resizebox{4.4cm}{!}{\includegraphics[scale=0.5]{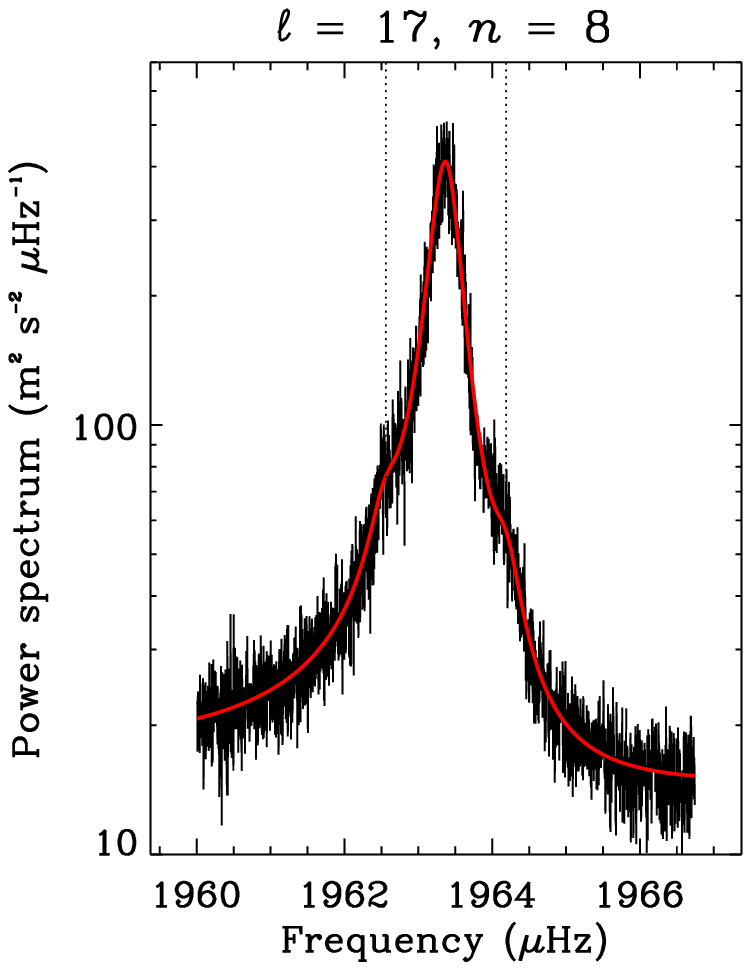}}
\end{tabular}
\caption{Examples of $m$-averaged power spectra in the case of the $l=17$ multiplet 
for four different radial orders $n=2$, 4, 6, and 8. The red lines represent 
the best MLE fits (Eq.~\ref{eq:mlemodel}), including the closest $\delta m =  \pm 2$ spatial 
leaks, whose positions are illustrated by the dotted lines.} 
\label{fig:mspec}
\end{center}
\end{figure*}

It can also be derived that the errors on the $a$-coefficients can be estimated as follows:
\begin{equation}
\sigma^{-2}_{a_i}=\frac{l^2}{2l+1}\Big(\sum_m[P_{l,m}^i(m/l)]^2\Big)\sigma^{-2}_{\nu_0},
\label{eq:a_error}
\end{equation}

\noindent
where $i$ is the $a$-coefficient order and $P_{l,m}^i$ the associated Clebsch-Gordan 
polynomials. The derivation of Eq.~\ref{eq:a_error} is detailed in Appendix~\ref{sec:a_errors}.

\subsubsection{$m/l$ dependence of the signal-to-noise ratio}
\begin{figure}[t]
\begin{center}
\includegraphics{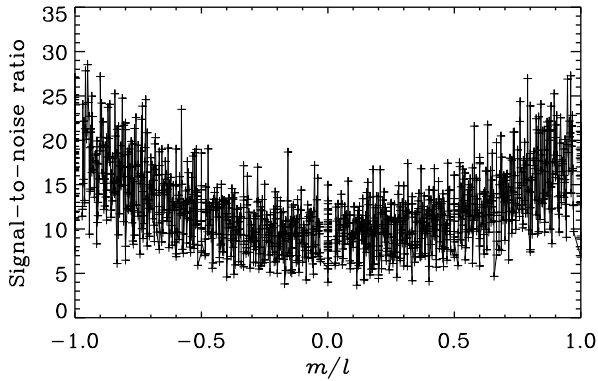}
\caption{Signal-to-noise ratio for modes up to $l=35$ as
a function of $m/l$, obtained with the 3960-day GONG dataset.} 
\label{fig:mdep}
\end{center}
\end{figure}

Figure~\ref{fig:mdep} shows the dependence in $m/l$ of the SNR in the GONG data. 
This was obtained with modes observed in the 3960-day GONG dataset below 2000~$\mu$Hz 
and of angular degree up to $l=35$. Both mode amplitude and background noise depends 
on the azimuthal order $m$ and can be described with polynomials with only even terms, 
the polynomials being different for both parameters. Note that any frequency dependence 
of the $m/l$ dependence is averaged out in Fig.~\ref{fig:mdep}.

The $m/l$ dependence of the SNR implies that the $a$-coefficients are not exactly 
orthogonal and that their errors are correlated (see Appendix~\ref{sec:a_errors}). However, as a first 
order approximation, the errors on the $a$-coefficients can be estimated by 
using Eq.~\ref{eq:a_error} (see Sec.~\ref{ssec:montecarlo}).

\subsubsection{Validation of the error estimates: Monte-Carlo simulations}
\label{ssec:montecarlo}
The formal uncertainties of the mode parameters and of the $a$-coefficients 
were verified through Monte-Carlo simulations. The artificial power spectra were 
simulated following the methodology described in \citet{fierry98}. In a first series 
of simulated spectra, the $m$ dependence in amplitude within a given multiplet ($n,l$) 
was introduced, the SNR being symmetric in $|m|$ around the $m=0$ spectrum. In a second 
series, no $m$ dependence was introduced, i.e., a constant SNR over $m$. The mean values 
of the formal errors returned by the MLE minimization were compared to the RMS value of 
the corresponding fitted parameter. The Monte-Carlo simulations showed that in both cases 
the formal uncertainties of the $m$-averaged spectra determined as in Sec~\ref{ssec:error} 
using a MLE minimization are a very good approximation of the errors.

\section{Comparison with other measurements}
\label{sec:compa}
\subsection{Comparison with spatially-resolved observations~($l\leq25$)}
GONG and MDI use two independent peak-finding approaches to extract the mode parameters. 
Developed in the early 1990s, and mostly unchanged since, they provide mode parameters on 
a routine basis. Time series of 108 days 
are used by the GONG project \citep{anderson90}, while the MDI project 
uses 72-day time series \citep{schou99}. Recently, \citet{korz05} 
developed a new and independent peak-finding method of the
individual-$m$ spectra, optimized to take advantage of the long,
spatially-resolved, helioseismic time series available today 
from both projects. 

\begin{figure*}[t]
\includegraphics[width=\textwidth]{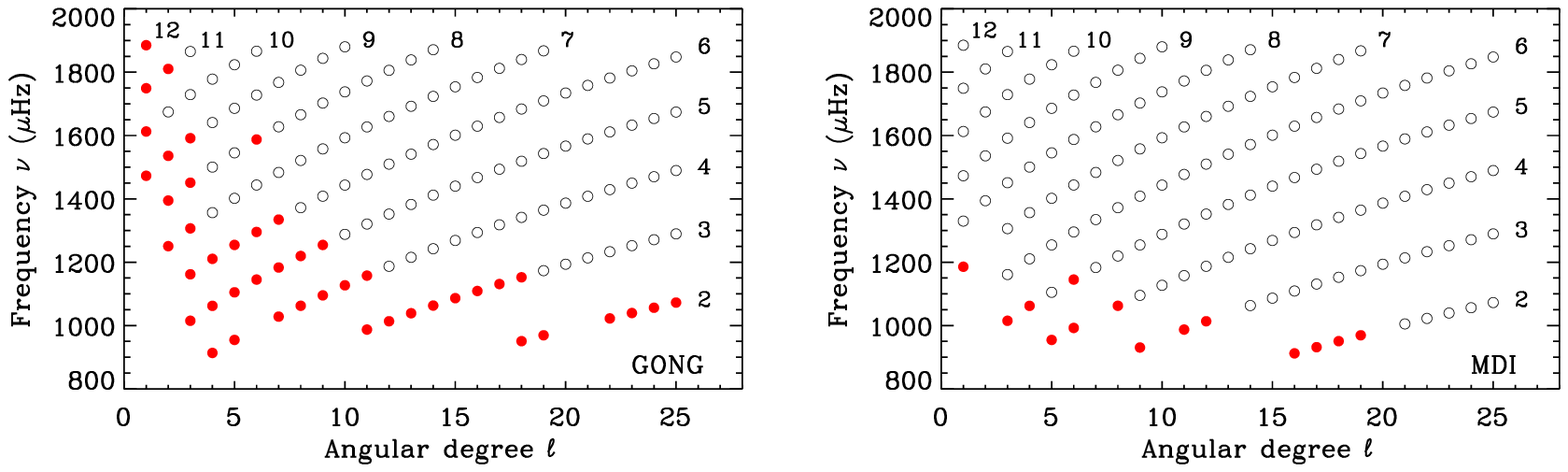}
\caption{$l - \nu$ diagrams of the low-frequency modes 
with $1 \leq l \leq 25$ measured in 2088 days of GONG ({\it left panel}) 
and MDI data ({\it right panel}). The open circles represent the modes 
measured by \citet{korz05} using a classic peak-fitting method of the individual-$m$ 
spectra, while the red dots correspond to the additional modes measured using 
the $m$-averaged spectrum technique that were not observed by \citet{korz05}. 
The ridges of same radial order are also indicated from $n=2$ to $n=12$.}
\label{fig:lnu2088d}
\end{figure*}

\subsubsection{Mode detection: $l-\nu$ diagram}
\citet{korz05} applied his peak-fitting to extract the low- and medium-degree ($l \leq 25$) 
mode parameters from both GONG and MDI observations using one 2088-day long time series, as 
well as using five overlapping segments of 728 days. In order to compare our results obtained 
with the $m$-averaged spectrum technique, we applied the procedure described in Sec.~\ref{sec:method}
to the same 2088 days of GONG and MDI observations (Table~\ref{table:series}). Figure~\ref{fig:lnu2088d} shows 
the $l-\nu$ diagrams of the low-frequency modes measured with the two different analyses in 
the case of the 2088-day GONG ({\it left panel}) and MDI ({\it right panel})  datasets. The modes 
measured by \citet{korz05} with a classic peak-fitting method of the individual-$m$ spectra are 
represented by the open circles. We considered that a given mode ($n,l$) from \citet{korz05} was 
detected when at least two of the $2l+1$ $m$ spectra were successfully fitted, which is enough to 
obtain estimates of the corresponding central frequency and first splitting coefficient $a_1$. 
The red dots represent modes measured with the $m$-averaged spectrum technique which were not 
observed by \citet{korz05}. A significantly larger number of low-frequency modes in the 2088-day 
GONG and MDI datasets (respectively 45 and 14 new modes) down to $\approx$~900~$\mu$Hz can be 
measured using the $m$-averaged spectrum technique. 

\subsubsection{Mode parameter and uncertainty comparisons}
\begin{figure*}[t]
\includegraphics[width=0.5\textwidth]{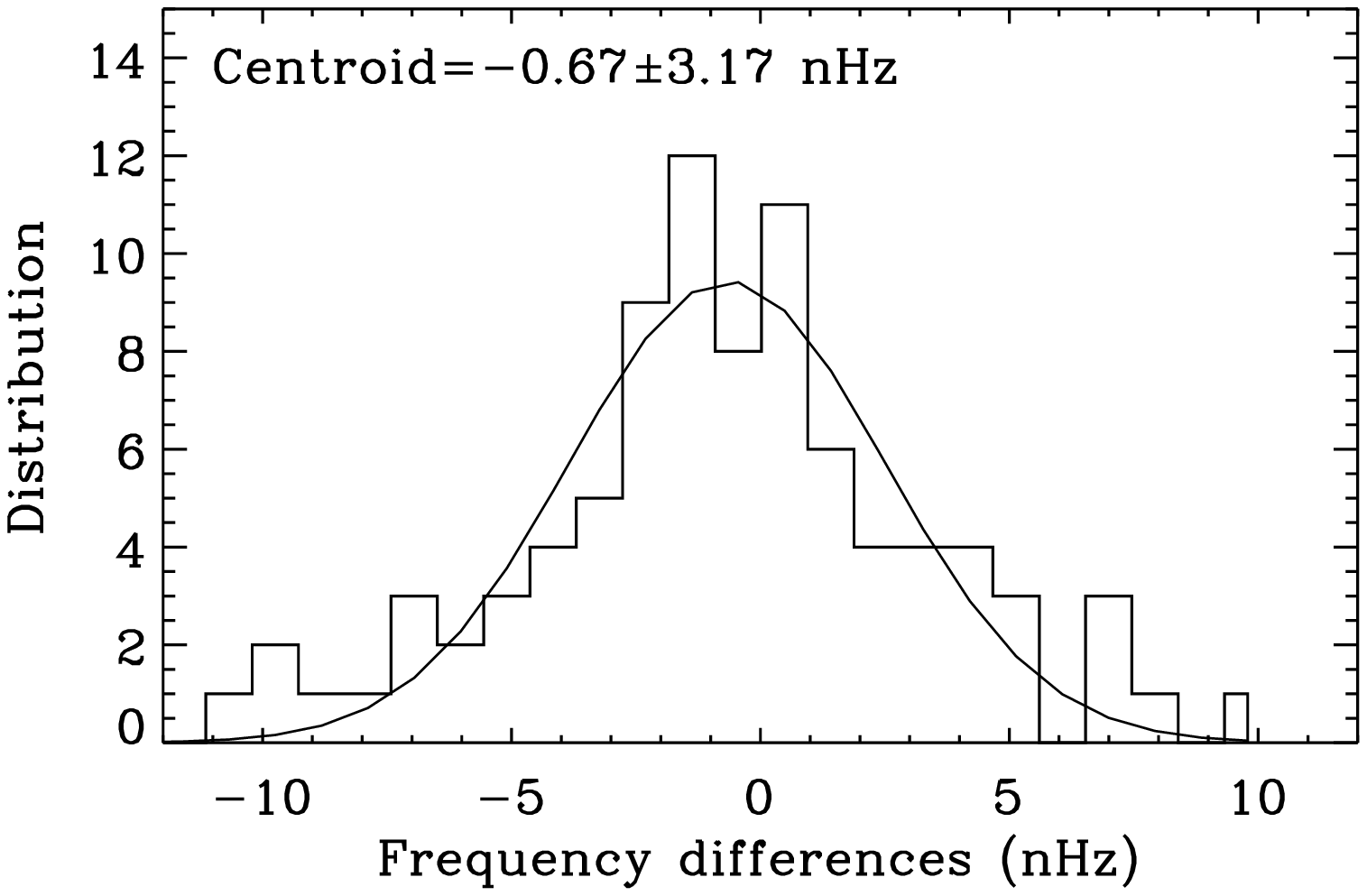}
\includegraphics[width=0.5\textwidth]{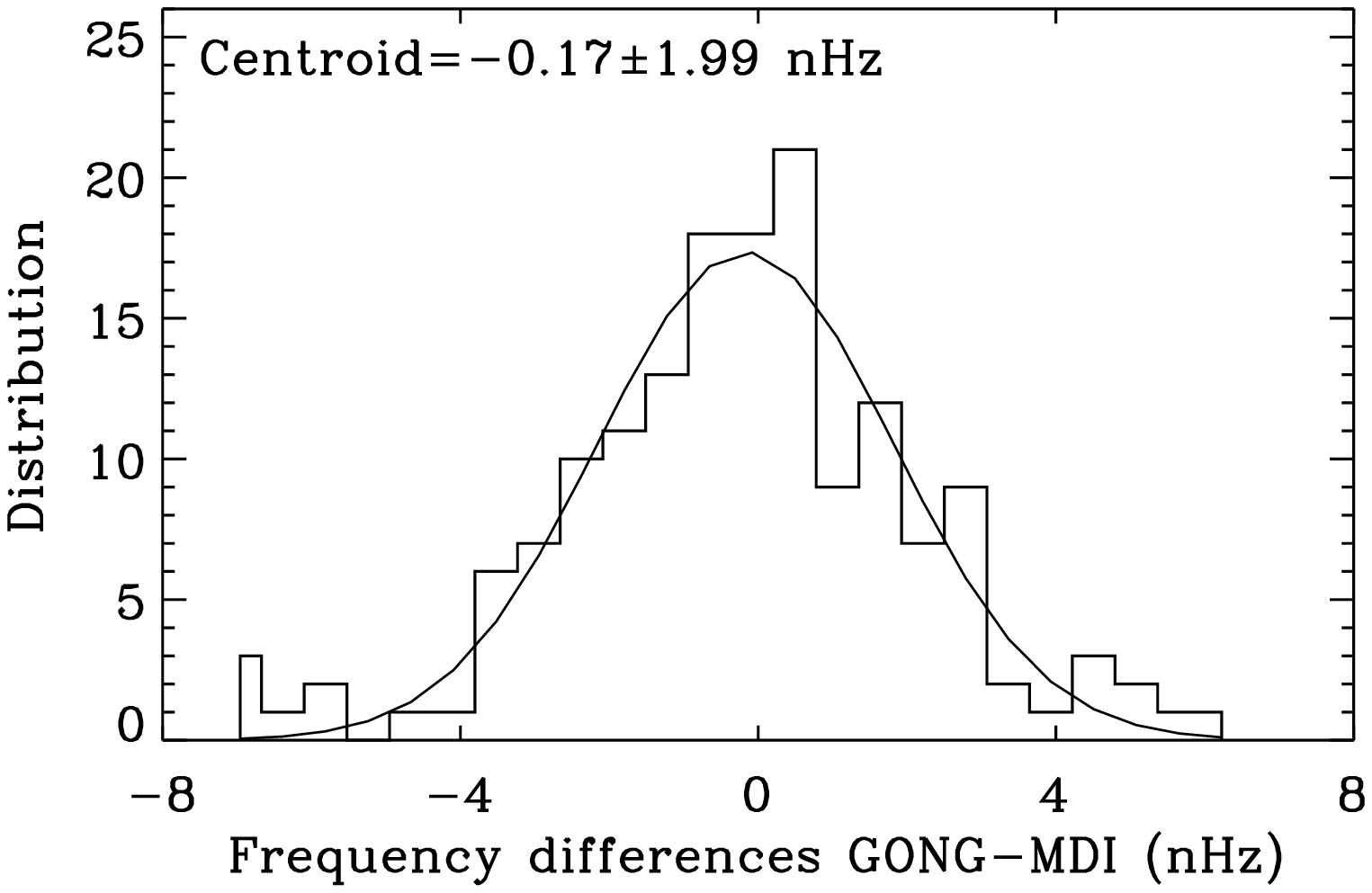}
\caption{{\it Left panel} - Histogram of the differences (in nHz) in the estimated central frequencies 
between the 2088-day GONG dataset using the $m$-averaged spectrum technique and the 
coveal 2088-day GONG \citet{korz05}'s estimates.  {\it Right panel} - Histogram of the 
differences (in nHz) in the estimated frequencies between the 2088-day GONG and MDI 
datasets using the $m$-averaged spectrum technique in the sense GONG minus 
MDI. The corresponding Gaussian function fits and their 
associated mean values and standard deviations are also indicated.}
\label{fig:hist2088d}
\end{figure*}

In order to check the accuracy of the technique and to identify any potential bias in 
our analysis, we compare the central frequencies and splitting coefficients obtained by 
the two methods. The individual-$m$ frequencies of \citet{korz05} were fitted using a 
Clebsch-Gordan polynomial expansion \citep{ritz91} in order to estimate the corresponding 
central frequencies and $a$-coefficients of each $(n,l$) multiplet. The formal uncertainties 
of the individual-$m$ frequencies were used as fitting weights. The left panel on 
Fig.~\ref{fig:hist2088d} shows the distribution of the differences in central frequencies 
below $\approx$~1800~$\mu$Hz of the common modes between the 2088-day GONG estimates measured using 
the $m$-averaged spectrum technique and from \citet{korz05} (as represented on Fig.~\ref{fig:lnu2088d}), 
demonstrating that there is no 
frequency dependence over the analyzed low-frequency range. 
The distribution was fitted by a Gaussian function, and its associated parameters (mean, 
standard deviation) are indicated on Fig.~\ref{fig:hist2088d}.
While, on average, the GONG central frequencies obtained using the $m$-averaged spectrum 
technique are less than 1~nHz smaller than \citet{korz05}'s estimates, this offset is not 
significant --- the corresponding standard deviation being about 5~times larger. The MDI
frequencies estimated with the $m$-averaged spectrum technique give comparable, insignificant 
mean differences with \citet{korz05}. Similar results are obtained with the splitting coefficients.

We also compared the low frequencies ($\nu \lesssim 1800\mu$Hz, see Fig.~\ref{fig:lnu2088d}) 
estimated in both the 
2088-day GONG and MDI datasets using the $m$-averaged spectrum technique. The right 
panel of Fig.~\ref{fig:hist2088d} represents the distribution of the frequency differences 
of the common modes, in the sense GONG minus MDI. The mean difference value is 
of -0.17~$\pm$~1.99~nHz, i.e. the GONG and MDI low-frequency modes are essentially 
the same. The mean difference in \citet{korz05}'s central frequencies between the 
2088-day GONG and MDI datasets for modes below 1800~$\mu$Hz is of 0.35~$\pm$~5.40~nHz. 
The splitting coefficient estimates are also consistent between the two datasets with in 
the case of the $a_1$ coefficient a mean difference of -0.04~$\pm$~0.31~nHz.

\begin{figure*}[t]
\includegraphics[width=0.5\textwidth]{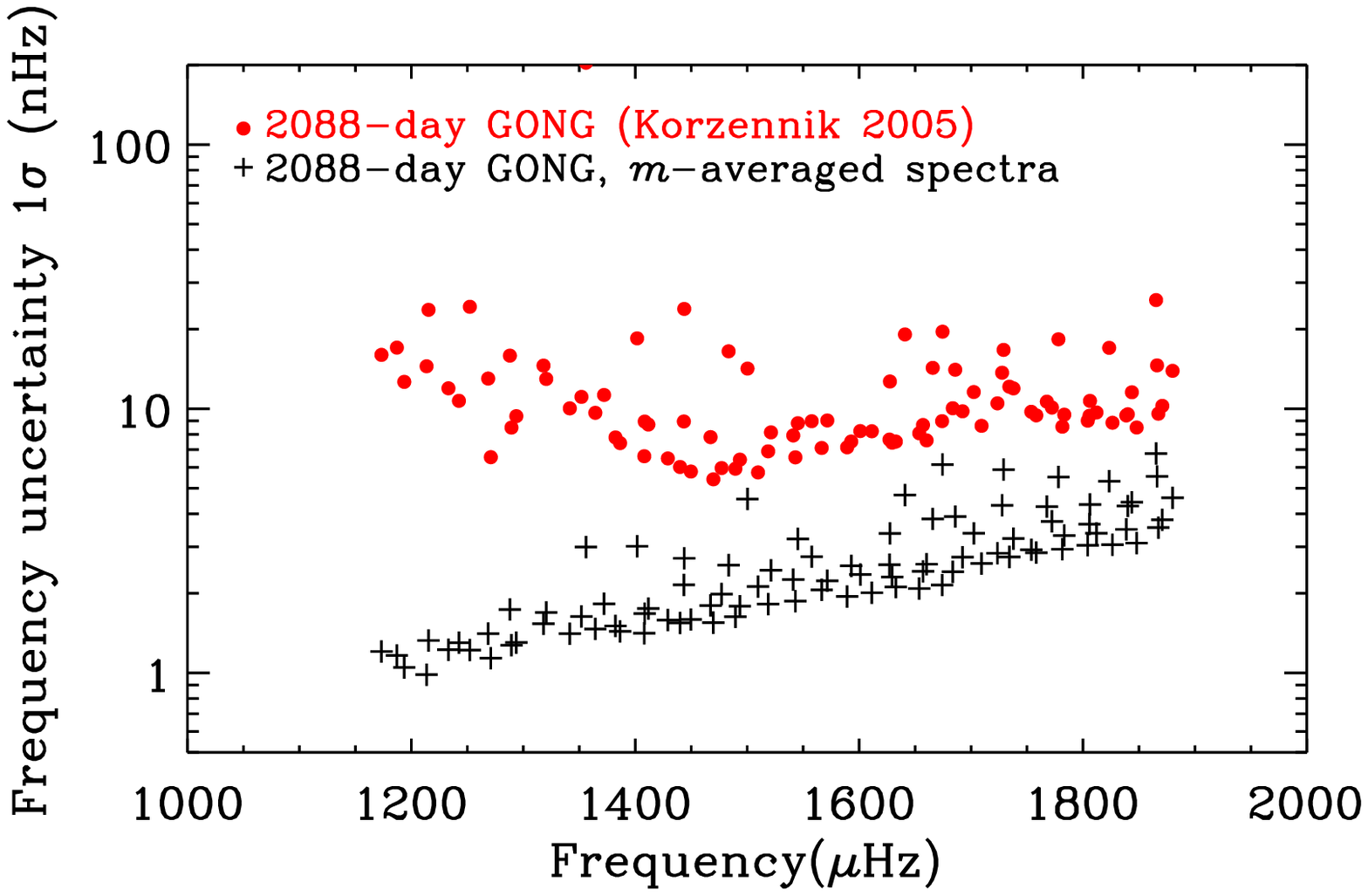}
\includegraphics[width=0.5\textwidth]{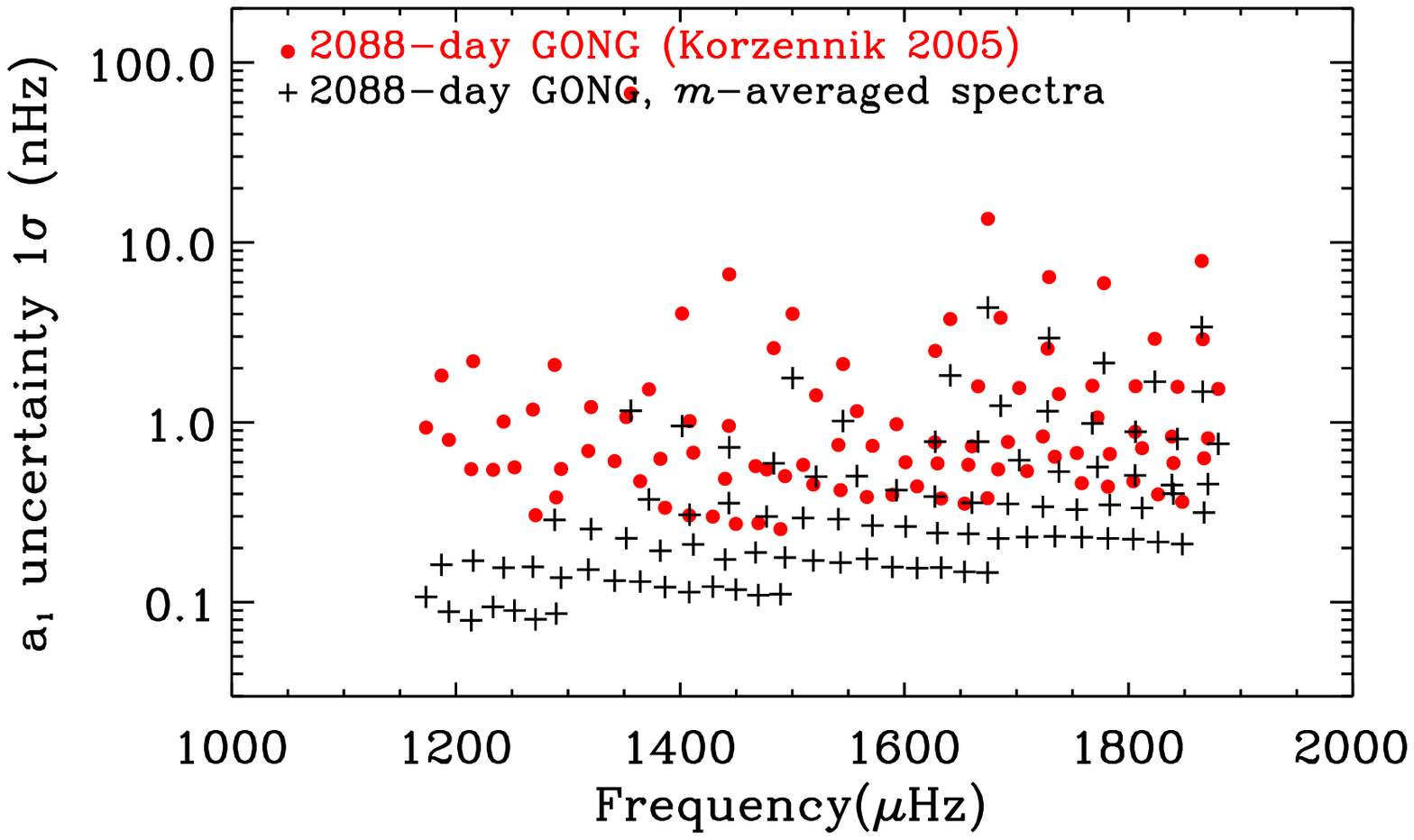}
\caption{Formal uncertainties (1~$\sigma$) in nHz of the central frequencies ({\it left panel}) 
and of the $a_1$ coefficients ({\it right panel}) as a function of frequency of the 
common modes measured in the 2088-day GONG dataset by \citet{korz05} fitting the 
individual-$m$ spectra (red dots) and by using the $m$-averaged spectrum technique 
(black plus signs).}
\label{fig:nu_2088d}
\end{figure*}

The left panel of Fig.~\ref{fig:nu_2088d} shows the formal $1\sigma$ uncertainties 
of the central frequencies of the measured common modes between the 2088-day GONG 
dataset analyzed in the present analysis and the coeval 2088-day GONG dataset 
from \citet{korz05} up to $\approx$~1800~$\mu$Hz. Our estimates of the frequency 
uncertainties are much smaller than those quoted by \citet{korz05}. However, Fig.~10 
in \citet{korz05} suggests that the errors are overestimated, and that his 
results ``might be too conservative". \citet{korz08} reported that in 
the case of a 2088-day long time series, as a first estimate, a multiplicative 
factor of 0.75 needs to be applied to the frequency uncertainties reported in \citet{korz05}. 
However, despite these uncertainty scaling issues, while the \citet{korz05}'s uncertainties 
show an increase with decreasing frequency from $\approx$~1500~$\mu$Hz, the uncertainties 
returned from the $m$-averaged spectrum technique do not show this increase, 
thanks to the higher SNR of the $m$-averaged spectrum than for the individual-$m$ spectra.

The uncertainties on the $a$-coefficients returned by the $m$-averaged spectrum technique 
are also smaller than the ones obtained by fitting the individual-$m$ spectra, as shown on 
the right panel of Fig.~\ref{fig:nu_2088d} in the case of the $a_1$ coefficients. As 
for the frequencies, the $a$-coefficients of the modes below $\approx$~1500~$\mu$Hz are 
better constrained using the $m$-averaged spectrum technique.

\subsection{Comparison with Sun-as-a-star observations~($l\leq3$)}
The spatially-resolved GONG and MDI instruments are not optimized to observe 
low-degree solar p modes below $l\leq3$, unlike the Sun-as-a-star, integrated-light 
instruments such as the space-based instrument GOLF onboard SOHO and the ground-based, 
multi-site BiSON network. The low-degree modes are of particular interest as they reach 
the very deep interior of the Sun. However, the spatially-resolved observations are 
still able to observe such low-degree oscillations.

\begin{figure*}[t]
\includegraphics[width=0.5\textwidth]{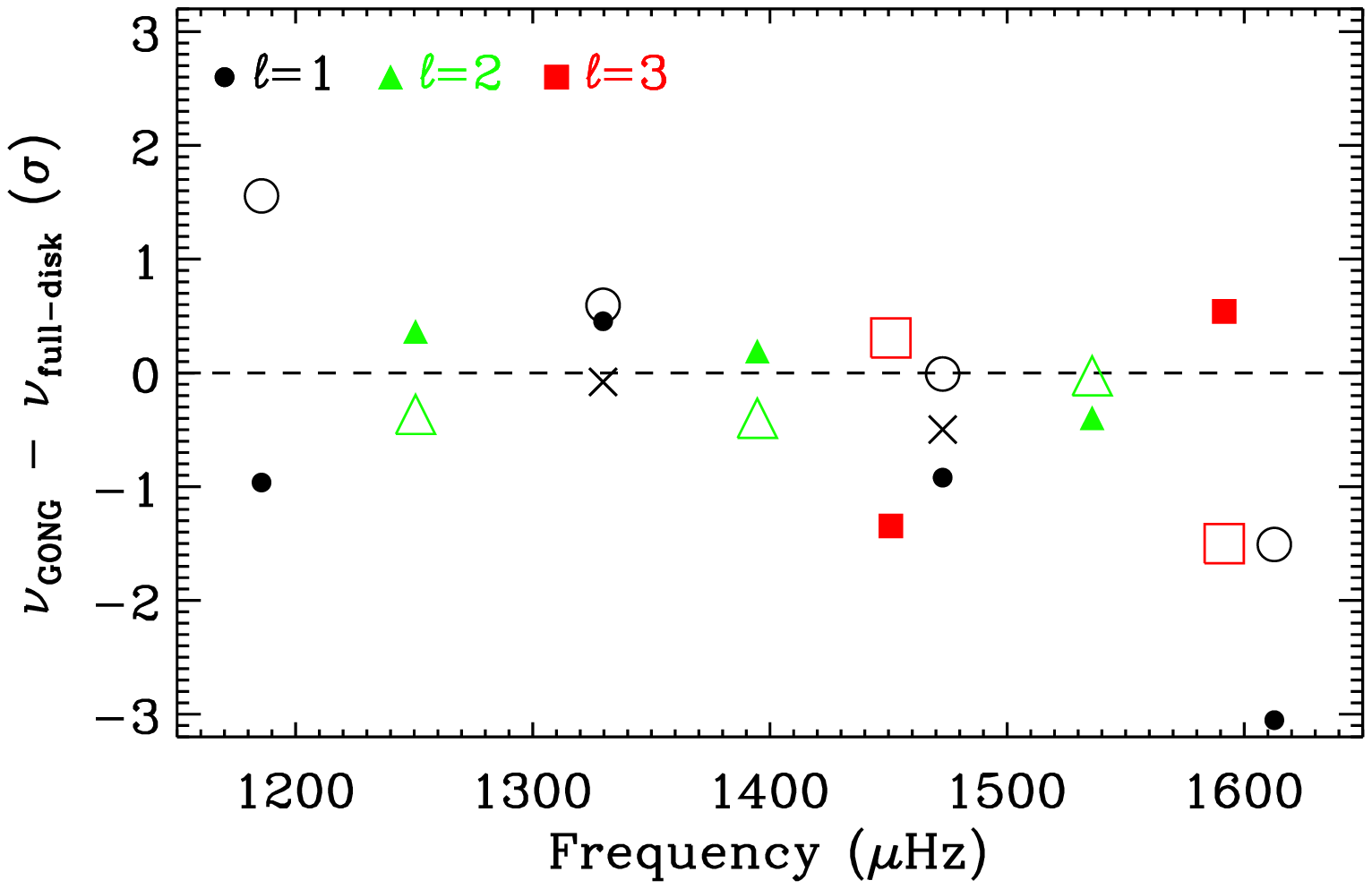}
\includegraphics[width=0.5\textwidth]{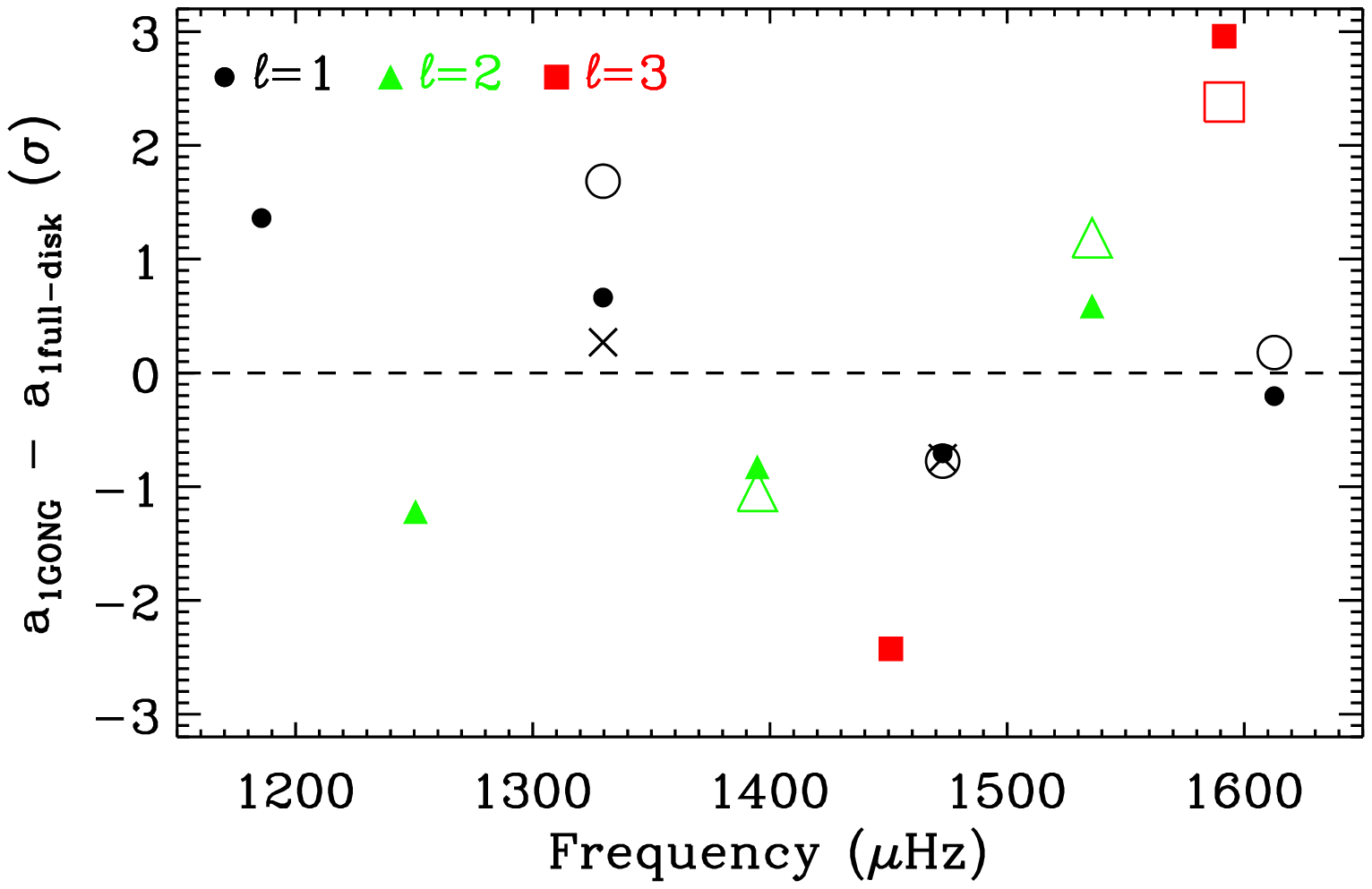}
\caption{{\it Left panel} - Frequency differences as a function of frequency of 
the common low-degree ($1\leq l \leq3$), low-frequency p modes observed 
in spatially-resolved GONG and MDI data using the $m$-averaged spectrum 
technique and in two Sun-as-a-star, full-disk observations, GOLF and BiSON, 
over comparable periods of time. The GOLF data were analyzed by two independent 
peak-fitting algorithms (R.~A Garc\'ia, filled symbols; P. Boumier, crosses). 
The comparison with BiSON is illustrated with the open symbols. The frequency 
differences were normalized by the combined errors. {\it Right panel} - Same as 
on the left panel, but for the splitting coefficient $a_1$.}
\label{fig:diff_fulldisk}
\end{figure*}

\begin{figure}[t]
\includegraphics[width=0.5\textwidth]{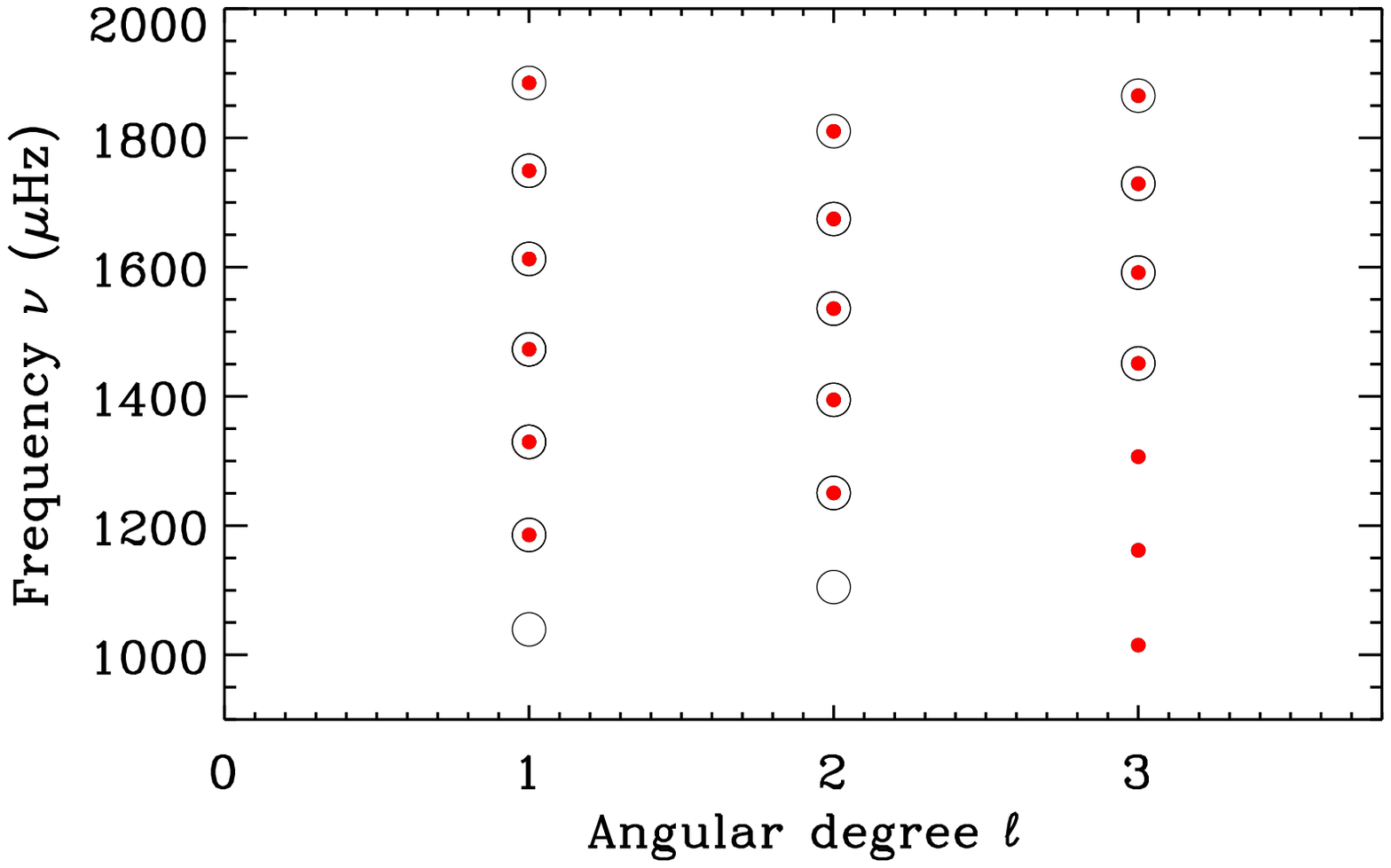}
\caption{$l-\nu$ diagram of the low-frequency, low-degree ($1\leq l \leq 3$) 
modes measured with the integrated-light instruments GOLF and BiSON (open circles) 
by R.~A Garc\'ia, P. Boumier, and W.~J. Chaplin (private communications)  and with 
the spatially-resolved instruments GONG and MDI using the $m$-averaged spectrum 
technique (red dots), over comparable periods of time.}
\label{fig:lnu_lowdeg}
\end{figure}

\begin{figure*}[t]
\includegraphics[width=\textwidth]{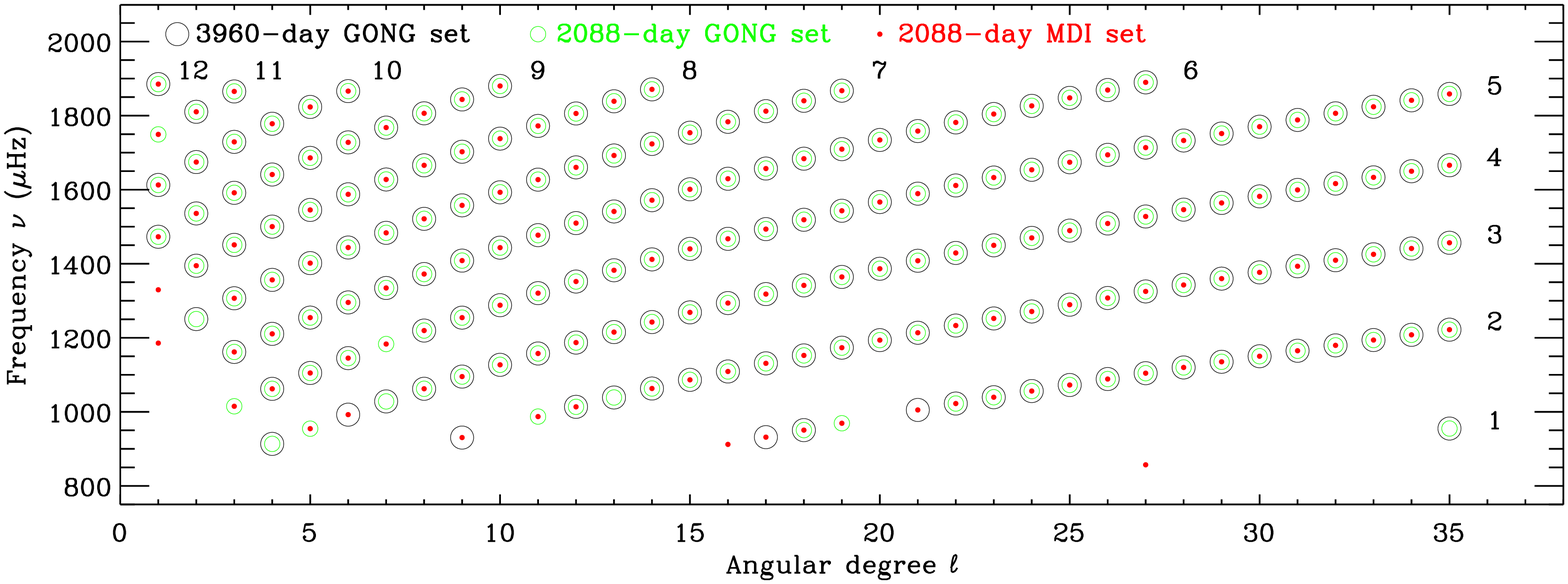}
\caption{$l - \nu$ diagram of the low-frequency solar p modes from $l=1$ up to $l=35$ observed
with the $m$-averaged spectrum technique in 3960 days of GONG observations (large black circles), 
and 2088 coeval days of GONG (medium green circles) and MDI (red dots) observations. The 
ridges of same radial order are also indicated from $n=1$ to $n=12$.}
\label{fig:odect}
\end{figure*}

Low-degree ($l \leq 3$) modes down to $\approx$~1000~$\mu$Hz are observed in 
both GONG and MDI data with the $m$-averaged spectrum technique as illustrated 
in Figs.~\ref{fig:lnu2088d} and \ref{fig:odect}. 
In order to test the capability and the precision of the $m$-averaged spectrum
technique to observe low-degree, low-frequency modes in spatially-resolved data, 
measurements obtained for $\approx$~11 years of the Sun-as-a-star GOLF and BiSON 
instruments were compared with the 3960-day GONG dataset and the 2088-day GONG and 
MDI datasets. The GOLF data were independently analyzed by two mode-fitting algorithms 
(R.~A. Garc\'ia, private communication\footnote{From 1996 April 11 to 2006 April 18.}, 
and P. Boumier, private communication\footnote{From 1996 April 11 to 2006 May 23.}). 
The BiSON observations come from a combination of frequencies obtained with different 
long time series analyzed in order to measure low-frequency modes \citep[W.~J. Chaplin, 
private communication. See also][]{chaplin02,broomhall07}. However, while imaging 
instruments give us access to all of the $2l+1$ individual-$m$ components,
only $l+1$ components can be clearly observed with integrated-sunlight technique. 
These various components have different spatial structure over the solar surface, 
which can lead to differences in the extracted central frequencies of the multiplet. 
\citet{chaplin04} and \citet{app07} derived expressions to predict the differences 
between the low-degree frequencies extracted from spatially-resolved (as MDI and GONG) 
and Sun-as-a-star (as GOLF and BiSON) observations. However, in the following, as a 
first approximation, we compared directly the extracted mode parameters.

The comparisons of the estimated mode frequencies and $a_1$ rotational splittings 
between the common low-degree ($1\leq l \leq3$), low-frequency modes in the two types 
of observation are shown on the left and right panels respectively of Fig.~\ref{fig:diff_fulldisk}. 
The three different datasets and analysis methods give consistent results, for both 
the frequency and the splitting coefficient $a_1$. 
Of course, this is only assuming that the different subsets of observed multiplets from 
both types of observational technique ``see" the same central frequencies. 

Thanks to decade-long available datasets, the low-degree, low-frequency modes are 
today measured lower than 1200~$\mu$Hz with high precision, demonstrated by the 
consistency in the extracted parameters from different instruments using distinct and 
independent analysis. Figure~\ref{fig:diff_fulldisk} also demonstrates that spatially-resolved 
observations can provide as accurate measurements of the low-degree modes as the 
Sun-as-a-star instruments do. Moreover, the $m$-averaged spectrum technique allows 
the observation of lower radial-order $l=3$ modes than the integrated-light GOLF and 
BiSON observations, for commensurate observation lengths, thanks to the observations of 
the $2l+1$ components (Fig.~\ref{fig:lnu_lowdeg}).  

\section{Mode parameters of the low-frequency oscillations}
\label{sec:data_results}
The $m$-averaged spectrum technique has been applied to 3960 days 
of GONG observations (see Sec.~\ref{sec:obs}), spanning most of the 
11 years of solar cycle 23. The analysis covered low-frequency modes 
with angular degrees from $l=1$ to $l=35$. Oscillation multiplets well below 
1000~$\mu$Hz were detected with good precision, such as the 
modes $l=4$, $n=4$ at $\approx$~913.5~$\mu$Hz; $l=9$, $n=3$ at $\approx$~930.5~$\mu$Hz;
$l=16$, $n=2$ at $\approx$~912.1~$\mu$Hz; or $l=31$, $n=1$ at $\approx$~907.5~$\mu$Hz. 
Some examples are illustrated in Figs.~\ref{fig:spec_l3} 
and \ref{fig:spec_l16}. These low horizontal-phase-velocity modes do not penetrate 
deeply into the Sun, but their very high inertias afford higher precision frequencies 
for the inversions. It is clear from Sec.~\ref{sec:compa} that this method allows us to 
observe modes that are otherwise lost in the background of each individual-$m$ spectrum 
of a given multiplet ($n,l$), and thus unobservable with a classic peak-fitting analysis. 
The $l-\nu$ diagram of the observed low-frequency modes ($1 \leq l \leq 35$) down to $n=1$ and
$\approx$~850~$\mu$Hz in the 3960-day GONG dataset and 2088-day GONG and MDI datasets 
with the $m$-averaged spectrum technique is shown on Fig.~\ref{fig:odect}.

\subsection{Mode linewidths, heights, and background levels}
\begin{figure*}[t]
\centering
\includegraphics[width=0.49\textwidth]{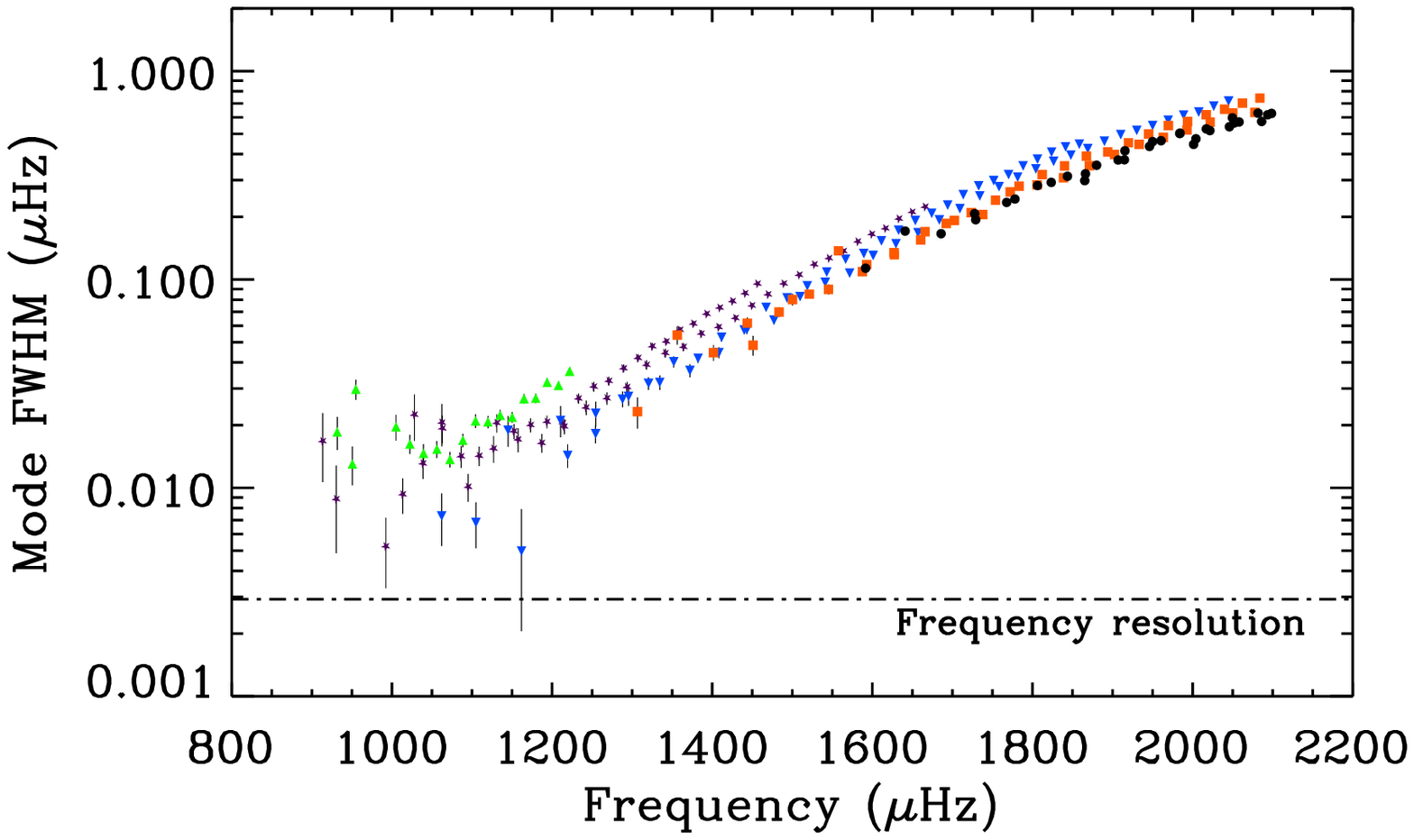} 
\includegraphics[width=0.49\textwidth]{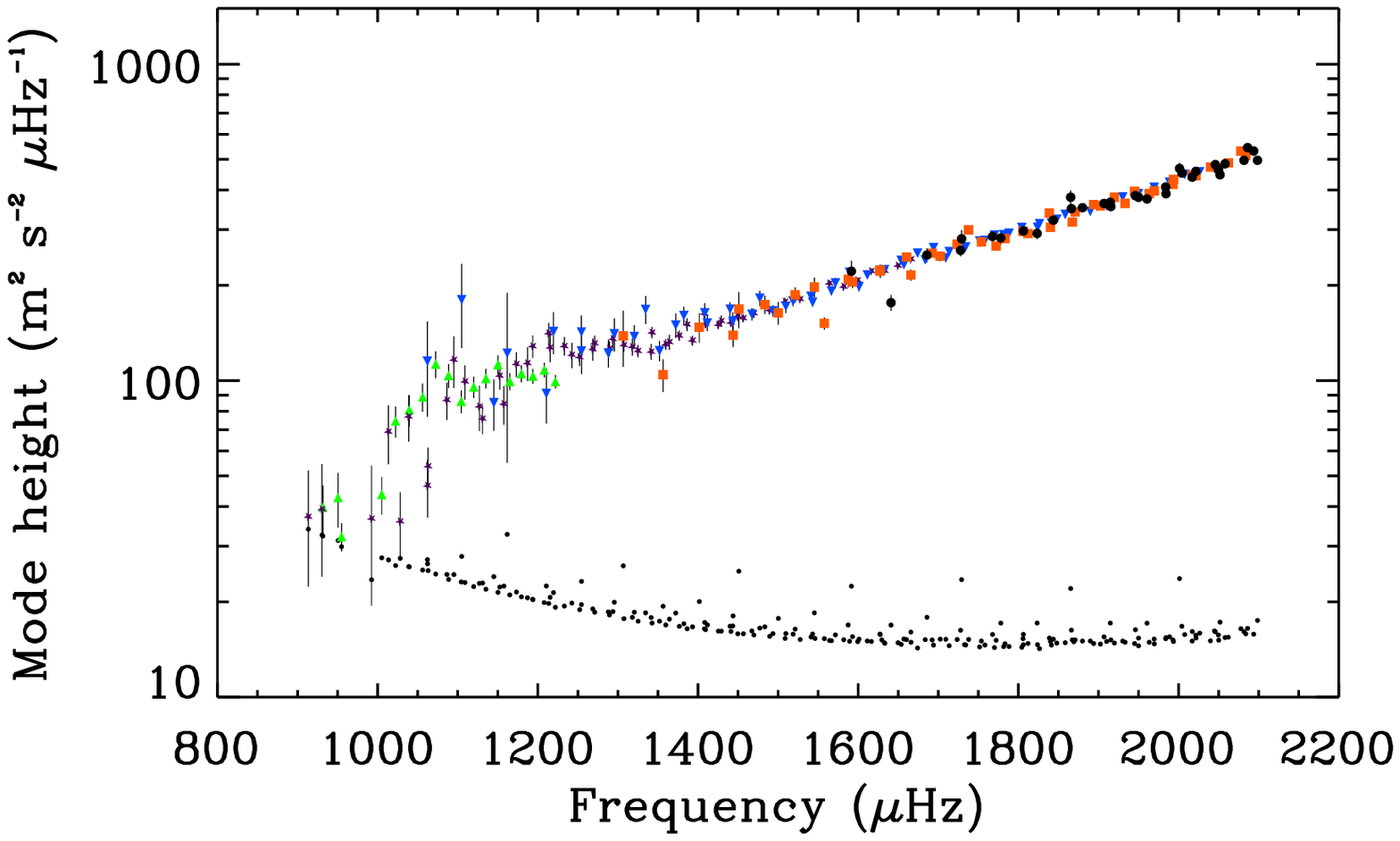} 
\includegraphics[width=0.49\textwidth]{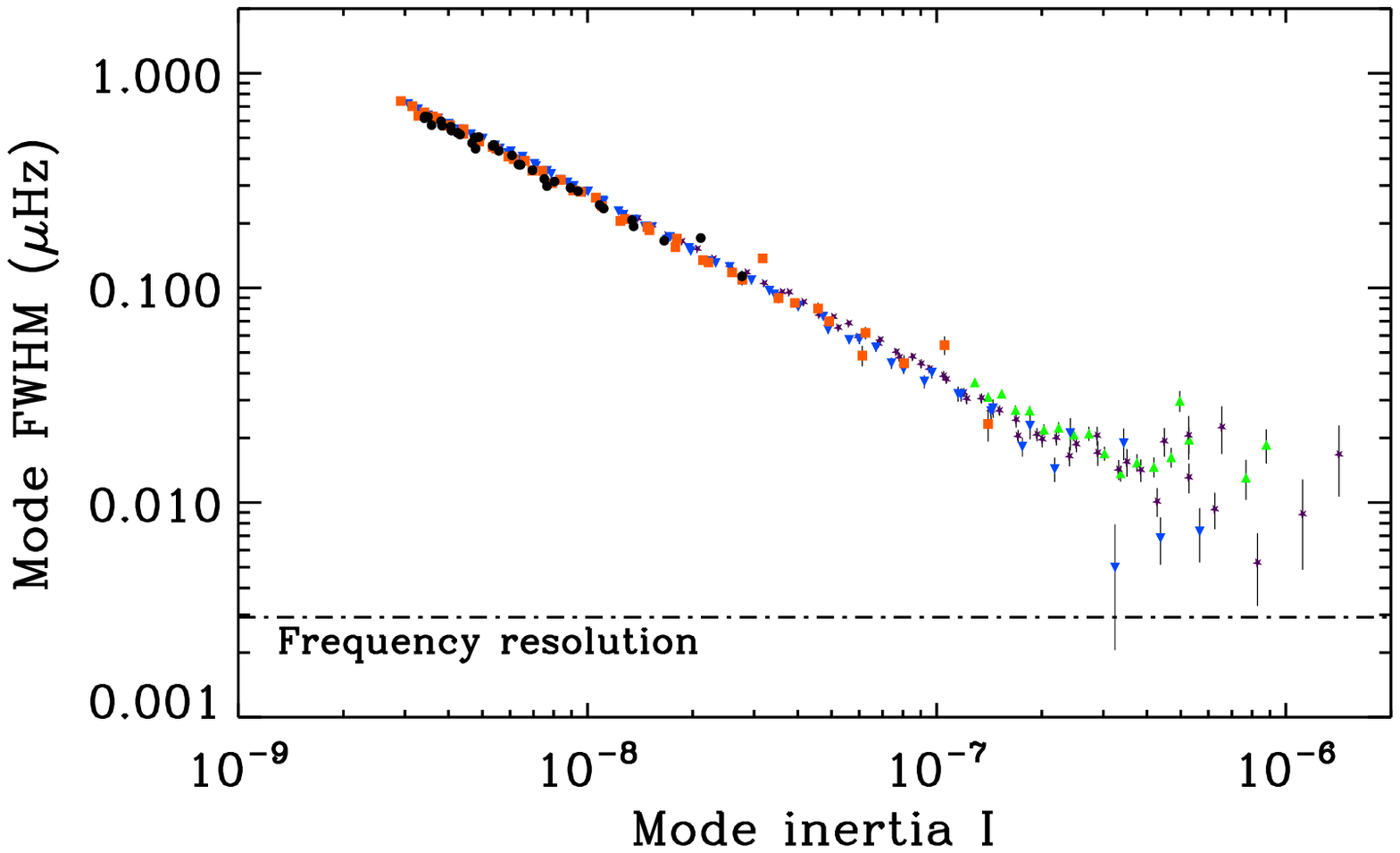} 
\includegraphics[width=0.49\textwidth]{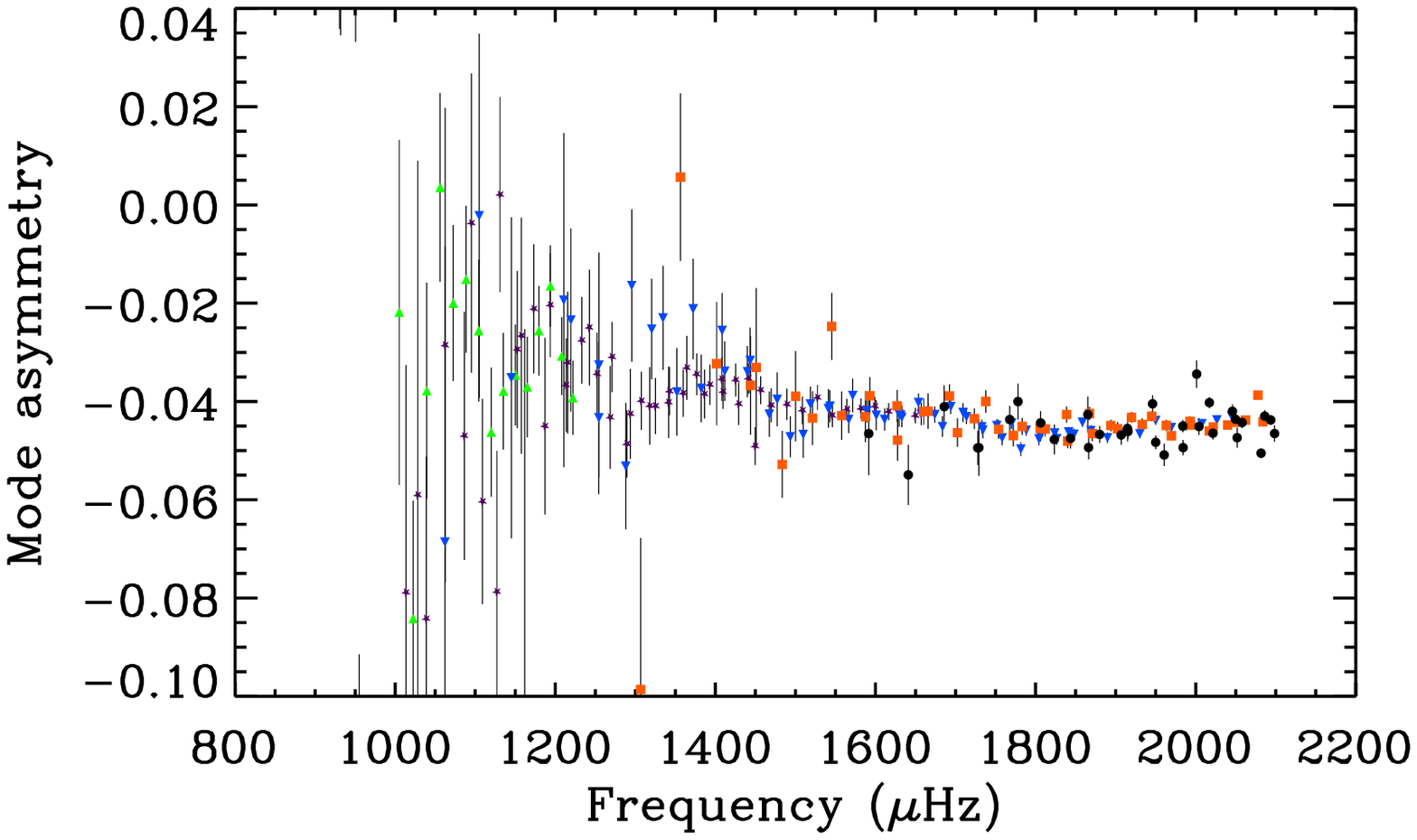} 
\caption{Mode \textsc{fwhm} $\Gamma_{n,l}$ ({\it upper-left panel}) and height $H_{n,l}$ 
({\it upper-right panel}) as a function of frequency for the low-frequency p modes $l \leq 35$ 
extracted from 3960 days of GONG data with the $m$-averaged spectrum technique. The different 
colors and symbols correspond to selected ranges of radial orders $n$: green triangles, 
modes with $n=1,2$; purple stars, $n=3,4$; blue upside-down triangles, $n=5,6$; 
orange squares, $n=7,8$; and black dots, $n\geq9$. The frequency resolution of the 
analyzed dataset is indicated on the upper-left hand-side plot, and the fitted 
background level ($B_{n,l}$) is also represented on the upper-right hand-side plot (small black dots).
{\it Lower-left panel} - Same as above, but for the mode \textsc{fwhm} $\Gamma_{n,l}$ 
as a function of the mode inertia ($I$), calculated from Christensen-Daslgaard model~S \citep{jcd96}. 
The frequency resolution is also indicated. {\it Lower-right panel} - Same as above, but for 
the mode asymmetry ($\alpha_{n,l}$).}
\label{fig:fwhm}
\end{figure*}

Figure~\ref{fig:fwhm} shows the fitted mode \textsc{fwhm}s $\Gamma_{n,l}$
({\it upper-left panel}) and mode heights $H_{n,l}$ ({\it upper-right panel}) of the 
measured low-frequency oscillations. The fitted background level is also 
represented on the right panel. The \textsc{fwhm}s and heights are extremely 
valuable tests of models of the physical processes responsible for the mode damping 
and excitation by the turbulent convective motions in the outer layers of the Sun: 
the mode damping is inversely related to the \textsc{fwhm} of the mode, and the mode
excitation is proportional to the mode height times the mode \textsc{fwhm} 
squared \citep[for a detailed description, see, e.g.,][]{salabert06}.
\colr{The leveling off of the mode widths observed below $\approx$~1100~$\mu$Hz, despite their dispersion becoming larger, could be a resolution effect, the peaks being then so 
narrow that the limiting resolution of the spectrum becomes an issue. Moreover, \citet{schou04} 
did not observe such behaviour at low frequency in MDI data with a 2952-day time series.}

As indicated by different colors and symbols on Fig.~\ref{fig:fwhm}, the fitted mode 
widths follow ridges for equal radial orders $n$. This dependence on angular degree ($l$) 
is directly related to the mode inertia ($I$) in terms of a power law, as illustrated on 
the lower-left panel of Fig.~\ref{fig:fwhm}. The $l$ dependence in the mode \textsc{fwhm}s 
is removed when represented as a function of the mode inertia $I$.

\begin{figure*}[t]
\centering
\includegraphics[width=\textwidth]{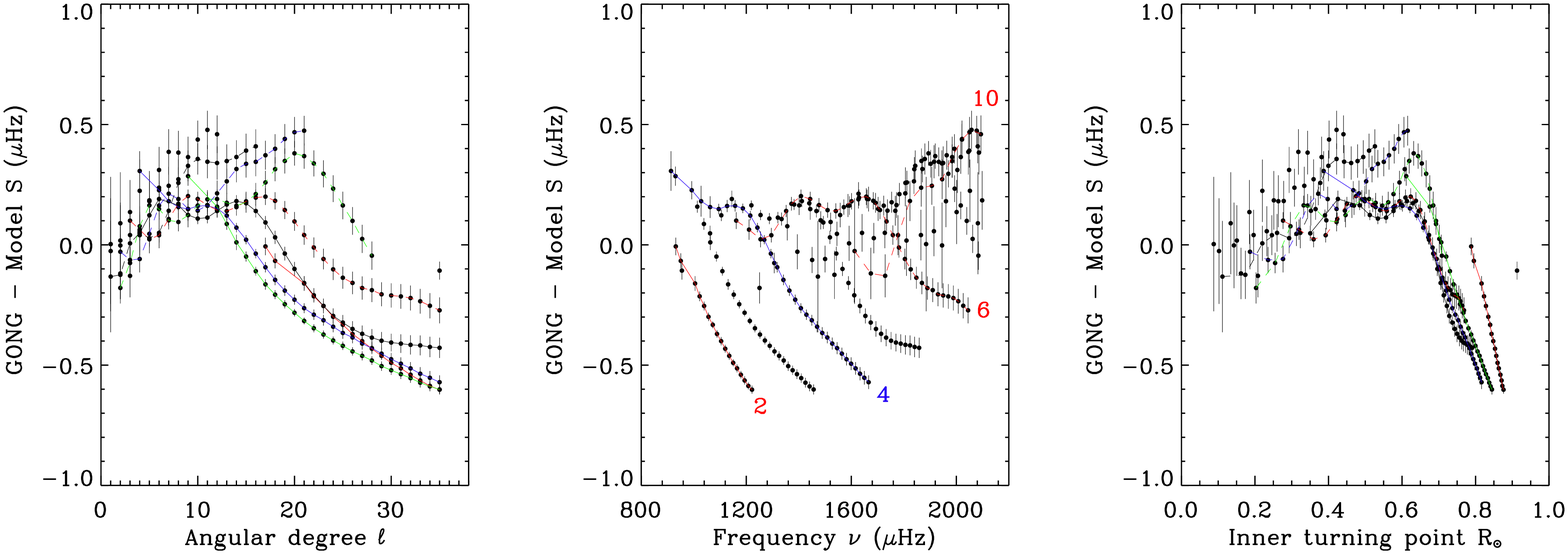}
\caption{Differences between the low-frequency central frequencies observed in 
3960 days of GONG data ($1 \leq l \leq 35$) using the $m$-averaged spectrum technique 
and the associated theoretical values, taken from Christensen-Dalsgaard's model~S \citep{jcd96}, 
as a function of the angular degree ({\it left panel}), of the frequency ({\it middle panel}),
and of the inner turning point ({\it right panel}). The uncertainties multiplied by a factor 20 
are also presented. Ridges of same radial orders are connected, and for guidance some of 
the corresponding radial orders are indicated in the middle panel.}
\label{fig:freq_th}
\end{figure*}

\subsection{Mode asymmetry}
The mode parameters extracted through the routine GONG and MDI 
peak-fitting pipelines are obtained by use of symmetric Lorenztian 
profiles (\citet{anderson90} and \citet{schou92} respectively). However, 
it was demonstrated that ignoring the peak asymmetry in the description of 
the acoustic modes leads to biais in the estimated mode parameters 
\citep[see Appendix~\ref{sec:impmodel} and][]{thiery00}. Today, most of the estimates of the mode asymmetries have 
been restricted to low degrees ($l \leq3$) only, from Sun-as-a-star, integrated-sunlight 
observations. However, \citet{korz05} used asymmetric profiles and presented estimates of 
the peak asymmetry for modes with angular degrees $1\leq l \leq25$, obtained with GONG and MDI 
observations. Recently, \citet{larson08} are planning to reprocess all the MDI medium-$l$ data 
including a set of corrections and improvements (such as the mode asymmetry) in the MDI pipeline 
algorithm itself.  

The asymmetry parameter ($\alpha_{n,l}$) in the low-frequency range, obtained by 
fitting the 3960-day GONG $m$-averaged spectrum ($1 \leq l \leq 35$), is shown on the 
lower-right panel of Fig.~\ref{fig:fwhm}. The extracted peak asymmetry is well constrained 
down to $\approx$~1400~$\mu$Hz, with a mean value of about $-0.044\pm0.002$, and no 
discernable $l$ dependence. The average asymmetry observed in the $m$-averaged spectrum 
is consistent with other measurements. For instance, the mean value observed by \citet{korz05} 
was about $-0.04$ for modes below 2000~$\mu$Hz and $l \leq 25$, once his estimates are transformed 
back into the \citet{nigam98}'s definition of the peak asymmetry. A comparable mean value 
is also observed at the lowest frequencies for which asymmetries were reported in Sun-as-a-star, 
integrated-sunlight observations \citep[e.g.,][]{thiery00}.

\subsection{Mode frequencies}
Figure~\ref{fig:freq_th} shows the frequency differences (in $\mu$Hz) between the 
fitted low-frequency modes observed in the 3960-day GONG dataset using the $m$-averaged 
spectrum technique and the corresponding theoretical values calculated from 
Christensen-Dalsgaard's model~S \citep{jcd96}. The corresponding frequency uncertainties were 
multiplied by 20 to render them visible. These comparisons are represented as a function of 
the angular degree ({\it left panel}), of the frequency ({\it middle panel}), and of the inner 
turning point ({\it right panel}). Modes of equal radial orders are connected.  As these 
differences between observed and theoretical frequencies show, there is still room to improve the 
model of solar internal structure. Note that the right panel on Fig.~\ref{fig:freq_th} illustrates 
also the wide range of depths of penetration that these low-frequency modes cover.

\begin{figure}[t]
\includegraphics[width=0.5\textwidth]{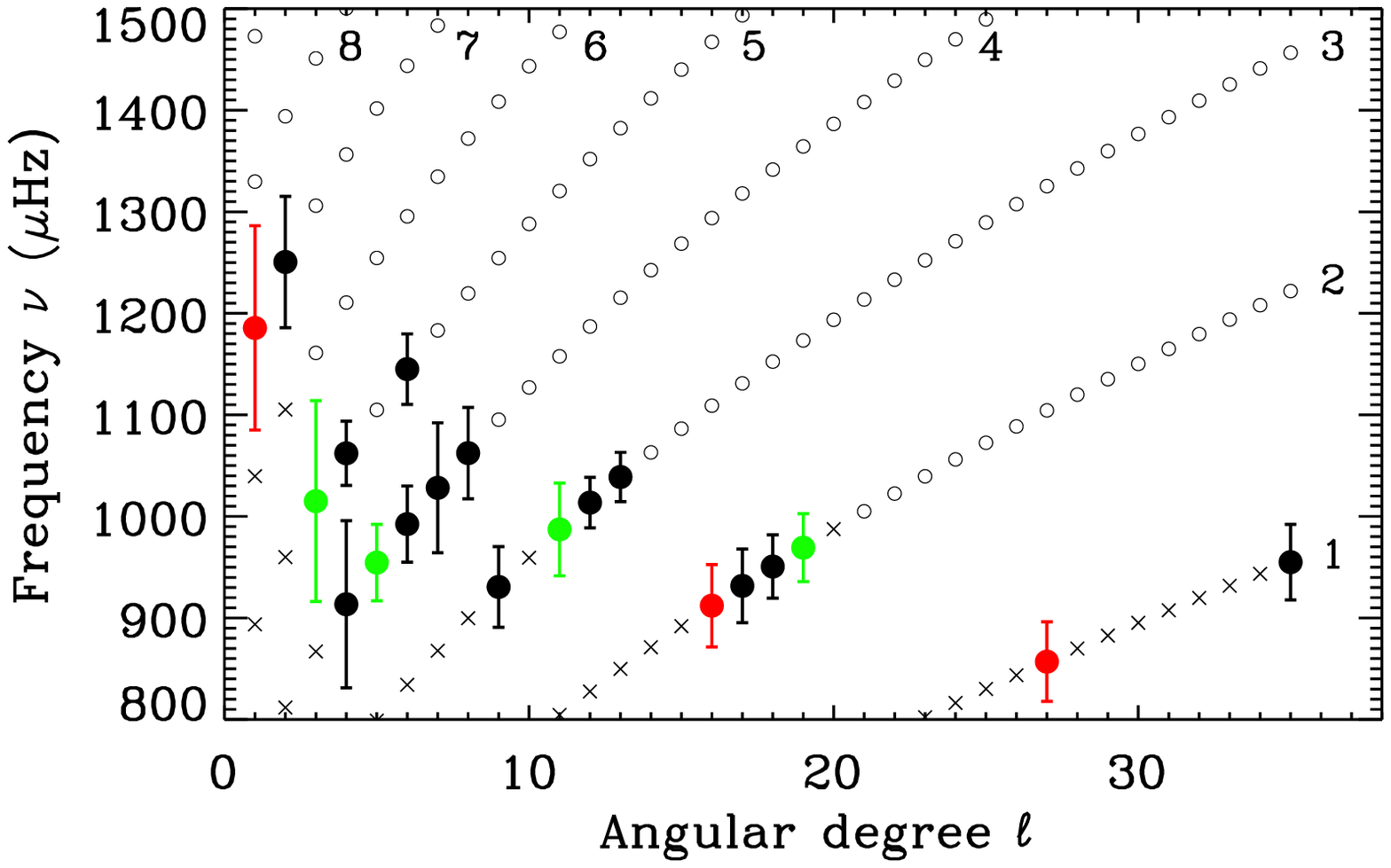}
\caption{$l - \nu$ diagram of the new low-frequency p modes observed in spatially-resolved 
data in the range of angular degrees $1\leq l \leq35$ (black dots: observed in the 3960-day 
GONG dataset / green dots: observed in the 2088-day GONG dataset / red dots: modes observed 
in the 2088-day MDI dataset). The corresponding frequency uncertainties were multiplied 
by $2\times10^4$. The already known modes are represented by the open circles, 
and the predicted modes by the crosses. The ridges of same radial order are also 
indicated from $n=1$ to $n=8$.}
\label{fig:newmodes}
\end{figure}

\section{Conclusion and discussion}
\label{sec:conc}
We presented here an adaptation of the rotation-corrected, $m$-averaged spectrum 
technique to observe low signal-to-noise-ratio, low-frequency solar p modes in
spatially-resolved helioseismic data. For a given multiplet ($n,l$), the shift 
coefficients describing the differential rotation- and structural-induced effects 
are chosen to \colr{maximize the likelihood} of the $m$-averaged spectra. The average of 
the $2l+1$ individual-$m$ spectra can result in a high signal-to-noise ratio when the 
individual-$m$ spectra have a too low signal-to-noise ratio to be successfully fitted. 
This technique was applied to long time series of the spatially-revolved GONG and MDI 
observations for low-frequency modes (i.e., approximately below 1800~$\mu$Hz) with low- and 
intermediate-angular degrees ($1\leq l \leq35$). We demonstrated that it allows us to measure 
lower frequency modes than with classic peak-fitting analysis of the individual-$m$ spectra. 
Figure~\ref{fig:newmodes} shows the new low-frequency solar p modes observed in spatially-resolved 
data using the $m$-averaged spectrum technique in long time series of both GONG and MDI observations.
Their central frequencies and splitting $a_1$ coefficients, as well as their associated uncertainties
are indicated in Table~\ref{table:newmodes}.
These normal modes of oscillation were predicted but were not measured previously. The potential of 
the $m$-averaged spectrum technique for increasing our knowledge of the solar interior is clearly 
illustrated on Fig.~\ref{fig:newmodes}, p modes well below 1000~$\mu$Hz being measured with a high 
accuracy \colr{thanks to their longer lifetimes}. 
We also demonstrated that the $m$-averaged spectrum 
technique returns unbiased results with no systematic differences with other long-duration measurements, 
which also include the asymmetry in the mode profile description. 

The oscillation parameters of these low signal-to-noise-ratio, low-frequency modes, such as their central 
frequencies, splittings, asymmetries, lifetimes, and heights were measured. 
These low-frequency p modes contribute to improve our resolution throughout the solar 
interior since they sample a large range of penetration depths. Moreover, because these modes 
have lower upper turning points in the outer part of the Sun, they are less sensitive to the 
turbulence and magnetic fields in the outer layers, which should make them extremely valuable 
for the study of the physical processes responsible for the oscillation excitation and damping 
by the turbulent convection.

\colr{We would like to recall that \citet{schou92}'s peak-finding approach consists in fitting the 
individual-$m$ spectra simultaneously by using a model in which the shift coefficients are introduced,
while in the present technique, the best shifts are determined first, based on the calculation of figure-of-merits (Sec.~\ref{ssec:det_acoef} 
and Appendix~\ref{sec:foms}), and then the rotation-corrected, $m$-averaged spectrum is fitted (Sec.~\ref{ssec:extraction}).}    

The development of the $m$-averaged spectrum technique towards both higher frequencies and 
larger angular degrees is one of the next step to be addressed, as also the analysis of 
shorter datasets, such as the canonical 108- and 72-day time series.

\begin{table}
\begin{center}
\caption{Set of the new low-frequency solar p modes observed in the GONG and MDI datasets 
with the $m$-averaged spectrum technique in the range $1\leq l \leq35$.}
\begin{tabular}{ccrr}
\hline\hline
\multicolumn{1}{c}{$l$} & \multicolumn{1}{c}{$n$}  & \multicolumn{1}{c}{Frequency} & \multicolumn{1}{c}{$a_1$-coefficient}\\ 
  &  & \multicolumn{1}{c}{($\mu$Hz)} & \multicolumn{1}{c}{(nHz)}  \\ \hline     
1  & 7 & 1185.599~$\pm$~0.005 & 431.491~$\pm$~6.161 \\
2  & 7 & 1250.555~$\pm$~0.003 & 428.263~$\pm$~2.286 \\
3  & 5 & 1015.046~$\pm$~0.005 & 430.154~$\pm$~2.471 \\
4  & 4 &  913.477~$\pm$~0.004 & 420.055~$\pm$~1.594 \\
4  & 5 & 1062.140~$\pm$~0.002 & 429.275~$\pm$~0.614 \\
5  & 4 &  954.560~$\pm$~0.002 & 430.712~$\pm$~0.596 \\ 
6  & 4 &  992.412~$\pm$~0.002 & 431.906~$\pm$~0.502 \\
6  & 5 & 1145.074~$\pm$~0.002 & 432.019~$\pm$~0.464 \\
7  & 4 & 1028.156~$\pm$~0.003 & 430.292~$\pm$~0.740 \\
8  & 4 & 1062.338~$\pm$~0.002 & 434.087~$\pm$~0.459 \\
9  & 3 &  930.540~$\pm$~0.002 & 430.045~$\pm$~0.363 \\ 
11 & 3 &  987.206~$\pm$~0.002 & 436.639~$\pm$~0.344 \\
12 & 3 & 1013.572~$\pm$~0.001 & 435.158~$\pm$~0.172 \\
13 & 3 & 1038.795~$\pm$~0.001 & 435.900~$\pm$~0.156 \\
16 & 2 &  912.080~$\pm$~0.002 & 436.331~$\pm$~0.213 \\
17 & 2 &  931.609~$\pm$~0.002 & 435.855~$\pm$~0.180 \\
18 & 2 &  950.625~$\pm$~0.002 & 436.627~$\pm$~0.146 \\
19 & 2 &  969.222~$\pm$~0.002 & 438.012~$\pm$~0.149 \\
27 & 1 &  856.964~$\pm$~0.002 & 437.741~$\pm$~0.123 \\
35 & 1 &  954.940~$\pm$~0.002 & 440.118~$\pm$~0.091 \\
\hline
\end{tabular}
\label{table:newmodes}
\end{center}
\end{table}

\acknowledgments
This work utilizes data obtained by the Global Oscillation Network Group (GONG) program, 
managed by the National Solar Observatory, which is operated by AURA, Inc. under a cooperative 
agreement with the National Science Foundation. The data were acquired by instruments operated 
by the Big Bear Solar Observatory, High Altitude Observatory, Learmonth Solar Observatory, 
Udaipur Solar Observatory, Instituto de
Astrof\'{\i}sica de Canarias, and Cerro Tololo Interamerican Observatory. The GOLF 
and MDI instruments onboard SOHO are cooperative efforts to whom we are indebted. 
SOHO is a project of international collaboration between ESA and NASA. 
BiSON is funded by the Science Technology and Facilities Council (STFC).
We thank the members of the BiSON team, and colleagues at the host institutes at each 
of the BiSON sites. The authors are particularly grateful to S.~G. Korzennik for providing 
us with the 2088-day MDI dataset, and to J. Schou for the MDI leakage matrix. The authors 
thank R.~A Garc\'ia, P. Boumier, and W.~J. Chaplin for providing estimates of GOLF and BiSON mode 
frequencies and $a_1$ rotational splittings observed in decade-long time series. The authors also 
thank S.~J. Jim\'enez-Reyes and J. Schou for their useful comments on the manuscript, \colr{and S.~G. Korzennik for helpful discussions during the various stages of this work.} D.~S. acknowledges 
the support of the NASA SEC GIP grant NAG5-11703. This work has been partially funded by the 
grant PNAyA2007-62650 of the Spanish National Research Plan.

\appendix

\section{Figures-of-merit and determination of the $a$-coefficients}
\label{sec:foms}
\begin{figure*}[t]
\includegraphics[width=1\textwidth]{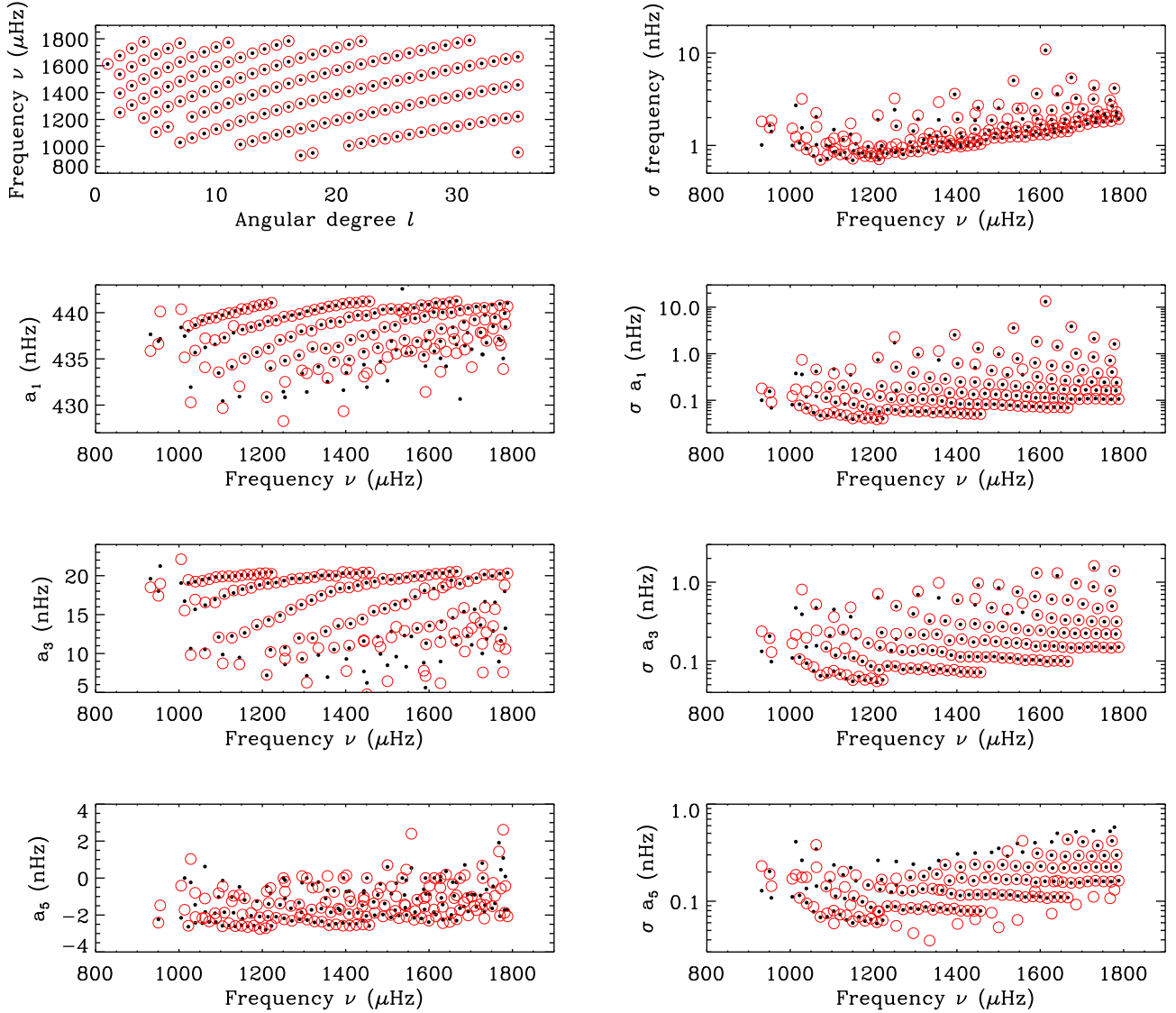}
\caption{{\it Left column} - $l-\nu$ diagram, and $a_1$, $a_3$, and $a_5$ splitting coefficients for 
the common, low-frequency p modes measured in the 3960-day GONG dataset
obtained by using the \colr{maximum likelihood} (circles) and the narrowest peak (dots) of the $m$-averaged
spectrum as figure-of-merit to determine the best estimates of the $a$-coefficients (see Sec.~\ref{ssec:det_acoef} 
and Fig.~\ref{fig:fom}). {\it Right column} - Associated $1\sigma$ formal uncertainties (nHz) of the mode central 
frequencies, and of the $a_1$, $a_3$, and $a_5$ splitting coefficients.} 
\label{fig:compfom}
\end{figure*}

The best estimates of the splitting $a$-coefficients are obtained by \colr{maximizing the likelihood} of the 
$m$-averaged spectrum (see Sec.~\ref{ssec:det_acoef}) through the calculation of a figure-of-merit (FOM). 
However, as shown on Fig.~\ref{fig:fom}, other criteria to define a FOM can be used, such as the 
narrowest peak (i.e., the minimum mode linewidth), or the minimum entropy of the resulting 
$m$-averaged spectrum. In order to compare the actual mode parameters and associated uncertainties 
obtained with two different definitions of the FOM, we applied the $m$-averaged spectrum technique to 
the 3960-day GONG dataset
by using both the \colr{maximum likelihood} and the narrowest peak in the $m$-averaged spectrum as FOM.  
Figure~\ref{fig:compfom} shows the corresponding central frequencies, and the odd $a_1$, $a_3$, 
and $a_5$ splitting coefficients of the common, measured low-frequency p modes. The associated formal uncertainties
are also represented.  The two FOMs return consistent mode parameters
within the error bars, the difference between the two being within the $3\sigma$ limit for all of the mode parameters.

\section{Impact of the fitting model used}
\label{sec:impmodel}
\begin{figure*}[t]
\includegraphics[width=0.5\textwidth]{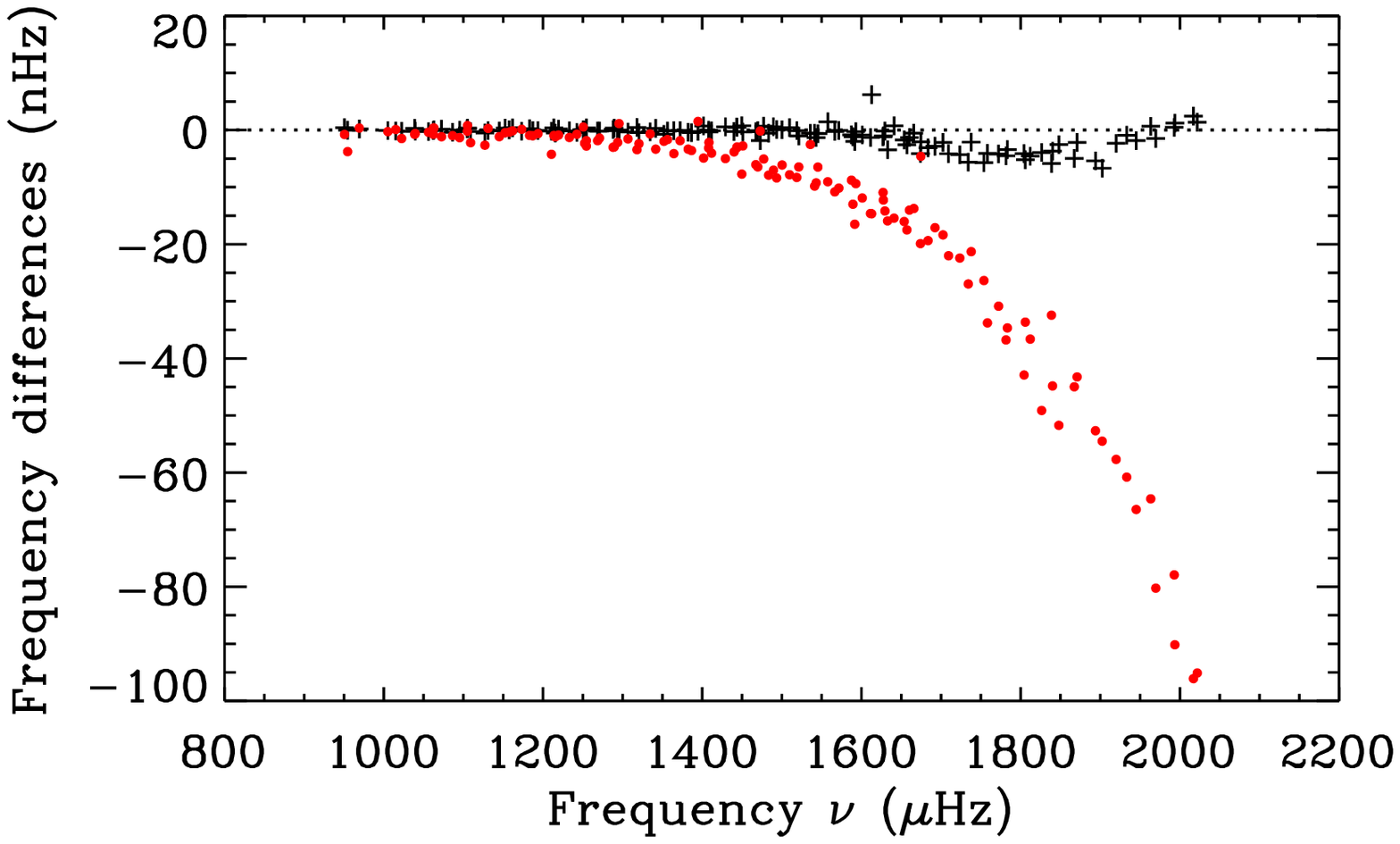}
\includegraphics[width=0.5\textwidth]{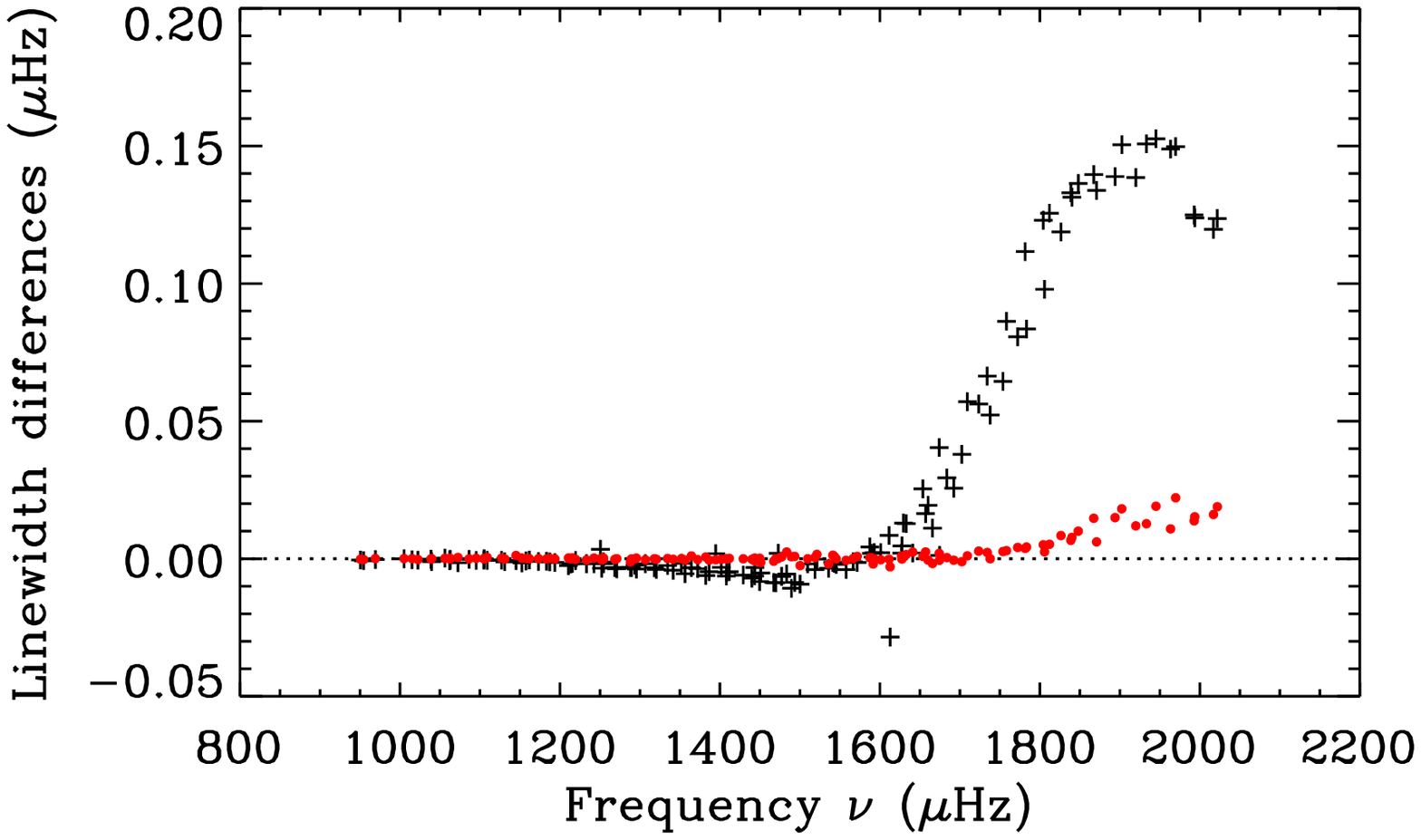}
\caption{Effect of asymmetry and spatial leaks on the fits. {\it Left panel} - Differences (in nHz) 
in the 2088-day GONG frequencies estimated using different models to describe the $m$-averaged spectrum: 
symmetric Lorentzian profile minus asymmetric Lorentzian profile (Eq.~\ref{eq:mlemodel}) both models 
including the first $\delta m \pm 2$ spatial leaks, i.e., {\em S2} minus {\em A2} (red dots); both 
asymmetric Lorentzian profiles (Eq.~\ref{eq:mlemodel}) but ignoring the closest $\delta m \pm 2$ spatial 
leaks for one of them, i.e., {\em A} minus {\em A2} (black plus signs). The differences in the extracted 
frequencies using different fitting profiles are represented as a function of 
frequency. {\it Right panel} - Same as the left panel, but for the differences (in $\mu$Hz) in the fitted mode linewidths.} 
\label{fig:nunoasym_2088d}
\end{figure*}

As a test of the dependence of the measured frequencies on the fitting model used to 
describe the $m$-averaged spectra, we fitted the $m$-averaged spectra using three different 
models: an asymmetric Lorenztian profile (Eq.~\ref{eq:mlemodel}) including the closest $\delta m \pm 2$ 
spatial leaks (hereafter called {\em A2} and used as the reference model); a symmetric Lorenztian profile 
including the $\delta m \pm 2$ spatial leaks (hereafter {\em S2}); and an asymmetric Lorenztian
profile (Eq.~\ref{eq:mlemodel}) but omitting the neighbouring $\delta m \pm 2$ spatial leaks 
(hereafter {\em A}). Figure~\ref{fig:nunoasym_2088d} shows the differences as a function of 
frequency in the 2088-day GONG low frequencies estimated using the $m$-averaged technique 
between {\em S2} and {\em A2} (red dots), and between {\em A} and {\em A2} (black plus signs), 
in both cases {\em A2} being the reference model. Ignoring the peak asymmetry in the fitting model 
leads to a systematic underestimation of the mode frequency as the frequency increases, the effect 
becoming particularly large above $\approx$~1400~$\mu$Hz (red dots). The differences become much 
larger than $3\sigma$, for example, at $\approx$~1800~$\mu$Hz, the fitted frequencies between {\em S2} 
and {\em A2} are about $20\sigma$ apart. These results obtained for modes below 2000~$\mu$Hz confirm 
previous observations, e.g. \citet{thiery00} who analyzed low-degree modes above 2000~$\mu$Hz 
in 805 days of GOLF data. 

On the other hand, omitting the spatial leaks has no effect below $\approx$~1600~$\mu$Hz, as 
they become well separated from the main peak because the corresponding mode linewidths are 
much smaller than their frequency separation. As the frequency increases, the mode linewidths 
increase, and ignoring the spatial leaks in the fitting model of the $m$-averaged spectrum between 
about 1600 and 2000~$\mu$Hz leads to an underestimation of the target mode frequency, the maximum 
difference occuring around 1800~$\mu$Hz. The frequency separation between the target mode and 
the $m$-leaks then becomes comparable to their linewidths and the lines blend together in 
the $m$-averaged spectrum. Above 2000~$\mu$Hz, this underestimation seems to vanish. Indeed, at 
that frequency range, the mode linewidths are much larger than the frequency separation, and the 
first spatial leaks (at least) are totally blended into the target mode in the $m$-averaged spectrum, 
having a much lower impact on the frequency determination. However, the effect of ignoring the peak 
asymmetry is much larger than that from ignoring the $m$ leaks even in the frequency range where 
the $m$ leaks have the strongest impact. For instance, at 1800~$\mu$Hz, the effect on the frequency 
underestimation by ignoring the mode asymmetry is about seven times larger than by ignoring the $m$ leaks.

As an example of the other mode parameters, the right panel of Fig.~\ref{fig:nunoasym_2088d} shows 
the differences in the extracted mode linewidths between the different fitting models. The color 
code is the same as for the differences in frequency represented on the left panel of 
Fig.~\ref{fig:nunoasym_2088d}. Ignoring the presence of the $m$ leaks in the fitting model leads 
to a 35\% overestimation at most of the extracted linewidths in the low-frequency range showing a 
maximum mismatch around 1900~$\mu$Hz. Interestingly, if the $m$ leaks are omitted, the linewidths 
are underestimated below $\approx$~1600~$\mu$Hz, showing a maximum 10\% underestimation around 
1500~$\mu$Hz. On the other hand, ignoring the peak asymmetry has a very small influence on the 
fitted linewidths in the low-frequency range. However, above $\approx$~1800~$\mu$Hz, the linewidths 
extracted using an asymmetric profile are systematically larger than the ones returned using a symmetric profile.  

\section{Derivation of errors from an $m$-averaged spectrum}
\label{sec:a_errors}
The derivation of the errors of the mode central frequencies and of the $a$-coefficients 
measured from the $m$-averaged spectrum technique is detailed here.

\subsection{Approximation of the statistics of the $m$-averaged spectrum} 
The $m$-averaged spectrum is obtained from the summation of $2l+1$ spectra assumed to be with $\chi^2$ 
with 2 d.o.f statistics each having a {\it different} mean or signal-to-noise ratio. All of the 
individual $m$-spectra are independent from each other. The solar background noise is assumed to 
depend on $m$ with a polynomial with only even terms (0, 2, etc...).  The amplitude of the modes 
is assumed to depend on $m$ with a different polynomial also with even terms (0, 2, etc...). 
In a first step, the $a_i$-coefficients are calculated to \colr{maximize the likelihood} of the resulting $m$-averaged spectrum.

Using \citet{app03}, we can derive an approximation of the statistics of the summation of 
the $2l+1$ spectra. The statistics of the $m$-averaged spectrum ${\cal S}$ can be approximated by a Gamma law given by:
\begin{equation}
	p({\cal S})=\frac{\lambda^{\nu_1}}{\Gamma(\nu_1)} {\cal S}^{\nu_1-1}e^{-\lambda {\cal S}}
	\label{approx1}
\end{equation}
The mean and $\sigma$ are given by:
\begin{equation}
	E[{\cal S}]=\frac{\nu_1}{\lambda}\,\,{\rm and}\,\,\sigma^2=\frac{\nu_1}{\lambda^2}
\end{equation}
$\lambda$ and $\nu_1$ are then derived from the mean and $\sigma$ as:
\begin{equation}
	\lambda=\frac{E[{\cal S}]}{\sigma^2}\,\,{\rm and}\,\,\nu_1=\frac{E[{\cal S}]^2}{\sigma^2}
	\label{approx2}
\end{equation}
In our case, the mean $E[{\cal S}]$ and $\sigma$ are given by:
\begin{equation}
	E[{\cal S}]=\sum_{m=-l}^{m= l} f_{m}\,\,{\rm and}\,\,\sigma=\sqrt{\sum_{m=-l}^{m=l} f_{m}^2}
	\label{approx3}
\end{equation}
where $f_m$ is the power spectrum for azimuthal order $m$ which can expressed as:
\begin{equation}
	f_m(\nu,\nu_0,a_i)=B_m(\nu)+A_m(\nu,\nu_0,a_i)
\end{equation}
where $\nu$ is the frequency, $\nu_0$ the central frequency , $a_i$ are the usual Ritzwoller-Lavely 
coefficients, $B_m$ is the background noise, $A_m$ is the profile of the mode (the linewidth and 
amplitude have been omitted for simplifying the notation).  We can write the noise as:
\begin{equation}
B_m(\nu)={\cal B}(\nu)(1+g_B(m))
\end{equation}
where the $m$-dependence is assumed to be independent of frequency. $g_B(m)$ is such that:
\begin{equation}
	E[B_m(\nu)]=(2l+1){\cal B}(\nu)
	\label{approx3}
\end{equation}
If the correction of the $a_i$ has been done properly, to the first order the 
$m$-averaged spectrum is independent of the $a_i$. We can write the mode amplitude as:
\begin{equation}
A_m(\nu)={\cal A}(\nu)(1+h_A(m))
\end{equation}
where the $m$-dependence is assumed to be independent of frequency.  $h_A(m)$ is such that
\begin{equation}
	E[A_m(\nu)]=(2l+1){\cal A}(\nu)
	\label{approx3}
\end{equation}
Then we find:
\begin{equation}
E[{\cal S}]=(2l+1)({\cal A}(\nu)+{\cal B}(\nu))
\end{equation}
and 
\begin{equation}
\sigma^2=(2l+1)\left[{\cal A}^2(\nu)\left(1+\alpha\right)+{\cal B}^2(\nu)\left(1+\beta\right)+2{\cal A}{\cal B}(\nu)\left(1+\rho\right)\right]
\end{equation}
with 
\begin{equation}
\alpha=\frac{1}{2l+1}\sum_{m=-l}^{m= l} h^2_A(m)
\end{equation}
and 
\begin{equation}
\beta=\frac{1}{2l+1}\sum_{m=-l}^{m= l} g^2_B(m)
\end{equation} 
\begin{equation}
\rho=\frac{1}{2l+1}\sum_{m=-l}^{m= l} h_A(m)g_B(m)
\end{equation}
we finally get for $\lambda$ and $\nu_1$ the following:
\begin{equation}
\lambda=\frac{{\cal A}(\nu)+{\cal B}(\nu)}{{\cal A}^2(\nu)\left(1+\alpha\right)+{\cal B}^2(\nu)\left(1+\beta\right)+2{\cal A}{\cal B}(\nu)\left(1+\rho\right)}
\end{equation}
and $\nu_1$ as:
\begin{equation}
\nu_1=\frac{(2l+1)({\cal A}(\nu)+{\cal B}(\nu))^2}{{\cal A}^2(\nu)\left(1+\alpha\right)+{\cal B}^2(\nu)\left(1+\beta\right)+2{\cal A}(\nu){\cal B}(\nu)\left(1+\rho\right)}
\end{equation}
After simplification we get:
\begin{equation}
\lambda=\frac{{\cal A}(\nu)+{\cal B}(\nu)}{({\cal A}(\nu)+{\cal B}(\nu))^2+\alpha {\cal A}^2(\nu)+\beta {\cal B}^2(\nu)+ 2 \rho {\cal A}(\nu){\cal B}(\nu)}
\end{equation}
and 
\begin{equation}
\nu_1=\frac{(2l+1)({\cal A}(\nu)+{\cal B}(\nu))^2}{({\cal A}(\nu)+{\cal B}(\nu))^2+\alpha {\cal A}^2(\nu)+\beta {\cal B}^2(\nu)+ 2 \rho {\cal A}(\nu){\cal B}(\nu)}
\end{equation}
Using the dependence observed in the GONG data, we have $\alpha \approx 0.17$, $\beta \approx 0.035$ and $\rho \approx 0.075$.  They are sufficiently small such that we have:
\begin{equation}
\lambda \approx \frac{1}{{\cal A}(\nu)+{\cal B}(\nu)}
\end{equation}
and 
\begin{equation}
\nu_1 \approx (2l+1)
\end{equation}
then we find the following statistics for the $m$-averaged spectrum:
\begin{equation}
	p({\cal S})=\frac{1}{{\Gamma(2l+1)}}\frac{{\cal S}^{2l}}{({\cal A}(\nu)+{\cal B}(\nu))^{2l+1}} e^{-\frac{{\cal S}}{ {\cal A}(\nu)+{\cal B}(\nu)}}
	\label{approx1}
\end{equation}
After a change of variable $u={\cal S}/(2l+1)$, we have
\begin{equation}
	p(u) \propto \frac{1}{({\cal A}(\nu)+{\cal B}(\nu))^{2l+1}} e^{-\frac{(2l+1)u}{ {\cal A}(\nu)+{\cal B}(\nu)}}
	\label{approx1}
\end{equation}
When we use MLE, we minimize the following
\begin{equation}
{\cal L}(\nu,\nu_0,a_i)=-\ln p(u) = -(2l+1) \left[\ln ({\cal A}(\nu)+{\cal B}(\nu))+\frac{u}{({\cal A}(\nu)+{\cal B}(\nu))}\right] +....
\label{eq:eq_mle}
\end{equation}
which shows that using the MLE applied to a $\chi^2$ with 2 d.o.f as prescribed by \citet{app03} 
is in the case of the $m$-averaged spectrum a good approximation.  It is not an approximation 
when averaging several power spectra of identical mean (or variance), i.e. when $\alpha=\beta=\rho=0$. 
Note that what we minimize is the sum over a range of frequency that can be approximated as:
\begin{equation}
L(\nu_0,a_i)=\int {\cal L}(\nu,\nu_0,a_i) {\rm d}\nu
\end{equation}

\subsection{Error bars on the central frequencies}
Error bars for frequency are derived from the inverse of the Hessian (second derivative of $L$) as:
\begin{equation}
\sigma^{-2}_{\nu_0}=\frac{\partial^2 {L}}{\partial \nu^2_0}.
\end{equation}
\citet{toutain94} showed that we could express the error bars as a function of the mode profile ${\cal P}$ (=${\cal A}+{\cal B}$) as:
\begin{equation}
\sigma^{-2}_{\nu_0}=T(2l+1)\int \frac{1}{{\cal P}^2(\nu)}\left(\frac{\partial{\cal P}}{\partial \nu_0}\right)^2 {\rm d}{\nu},
\label{eq:eq26}
\end{equation}
\noindent
where $T$ is the observation time. The $2l+1$ factor is due to the fact that the likelihood is $2l+1$ 
times larger than the likelihood of \citet{toutain94} (cf Eq.~23). Eq.~(\ref{eq:eq26}) 
shows that the error bars on the frequencies in the $m$-averaged spectrum will be $\sqrt{2l+1}$ smaller 
than for the {\it mean} of the individual modes. In deriving Eq.~(\ref{eq:eq26}), we assumed 
that $\langle u \rangle={\cal P}$.  This is an approximation good enough for getting the error 
bars on the frequency but not on the $a_i$.

\subsection{Error bars on the $a$-coefficients}
The error bars on the $a$-coefficients are derived from the inverse of the Hessian (second derivative of $L$) as:
\begin{equation}
h_{ij}=\frac{\partial^2 {L}}{{\partial a_i}{\partial a_j}}.
\end{equation}

As shown by \citet{toutain94}, these coefficients can be related to the mode profile as using Eq.~(\ref{eq:eq_mle}):
\begin{equation}
h_{ij}=T \sum_m \int  \frac{1}{{\cal P}_m^2(\nu)}\frac{\partial {\cal P}_m}{{\partial a_i}}\frac{\partial {\cal P}_m}{{\partial a_j}} {\rm d}\nu,
\end{equation}
using the following property:
\begin{equation}
\frac{\partial {\cal P}_m}{{\partial a_i}} =\frac{\partial {\cal P}_m}{{\partial \nu_0}} l P_{l,m}^i(m/l),
\end{equation}
where $P_{l,m}^i(m/l)$ are the Ritzwoller-Lavely polynomials. And we finally get:
\begin{equation}
h_{ij}=T l^2 \sum_m P_{l,m}^i(m/l)P_{l,m}^j(m/l) \int  \frac{1}{{\cal P}_m^2(\nu)}\left(\frac{\partial {\cal P}_m}{{\partial \nu_0}} \right)^2 {\rm d}\nu.
\label{eq:eq_32}
\end{equation}
We recognize the error bars of the frequency for the $m$ spectrum depending on the inverse 
of the signal-to-noise ratio $\beta_m$ \citep[as in][]{lib92}:
\begin{equation}
\sigma^{-2}_m=T \int  \frac{1}{{\cal P}_m^2(\nu)}\left(\frac{\partial {\cal P}_m}{{\partial \nu_0}} \right)^2 {\rm d}\nu.
\end{equation}
Finally Eq.~(\ref{eq:eq_32}) can be written as:
\begin{equation}
h_{ij}=l^2 \left(\sum_m P_{l,m}^i(m/l)P_{l,m}^j(m/l)\right) \sigma_m^{-2}.
\end{equation}
The errors $\sigma_{a_1}$ scale like $l^{-\frac{3}{2}}$ \citep[as in][]{veitzer93}. 
If the SNR is the same for all $m$, then we have $\sigma_{\nu_0}^{-2}=(2l+1)\sigma_{m}^{-2}$.  Thus, 
by simply using the orthogonality property of the $P_{l,m}^i$ polynomials, and as given 
in Sec.~\ref{ssec:error} (Eq.~\ref{eq:a_error}), we obtain the following expression to calculate 
the error bars of the $a$-coefficients in the $m$-averaged spectrum:
\begin{equation}
\sigma_{a_i}^{-2}=\frac{l^2}{2l+1} \left(\sum_m \left[P_{l,m}^i(m/l)\right]^2\right) \sigma^{-2}_{\nu_0}.
\end{equation}
All terms off of the diagonal are zero. Of course, when the SNR varies with $m$, the off-diagonal 
terms are non-zero and correlations appear.


\clearpage

\end{document}